\documentclass[table]{article}

\usepackage{listings}
\usepackage[utf8]{inputenc}
\usepackage[T1]{fontenc}
\usepackage{lmodern}
\usepackage{textcomp}
\usepackage{color}
\usepackage{enumerate}
\usepackage{graphicx}
\usepackage{grffile}
\usepackage{wrapfig}
\usepackage{rotating}
\usepackage{longtable}
\usepackage{multirow}
\usepackage{multicol}
\usepackage{changes}
\usepackage{pdflscape}
\usepackage{geometry}
\usepackage[normalem]{ulem}
\usepackage{amssymb}
\usepackage{amsmath}
\usepackage{amsfonts}
\usepackage{dsfont}
\usepackage{array}
\usepackage{ifthen}
\usepackage{hyperref}
\usepackage{natbib}
\usepackage{epstopdf} 
\usepackage{caption}
\usepackage[labelformat=simple]{subcaption}
\usepackage{booktabs}
\usepackage{siunitx} 
\usepackage{algorithm2e}
\usepackage{xr} 
\usepackage{amsthm,dsfont,amsmath}

\newcommand{\X}{X}
\newcommand{\Vn}{\mathbf{n}}
\newcommand{\VX}{\boldsymbol{X}}

\newcommand{\Y}{Y}
\newcommand{\VY}{\boldsymbol{Y}}

\newcommand{\Veta}{\boldsymbol{\eta}}
\newcommand{\Vvarepsilon}{\boldsymbol{\varepsilon}}
\newcommand{\Vmu}{\boldsymbol{\mu}}
\newcommand{\Vxi}{\boldsymbol{\xi}}
\newcommand{\set}{\mathcal{S}}
\newcommand{\param}{\theta}
\newcommand{\paramHat}{\hat{\theta}}
\newcommand{\Vparam}{\boldsymbol{\param}}
\newcommand{\VparamHat}{\boldsymbol{\paramHat}}

\newcommand\Information{\mathcal{I}}
\newcommand\Score{\mathcal{U}}

\newcommand\Real{\mathbb{R}}
\newcommand\half{\frac{1}{2}}
\newcommand\dpartial[2]{\frac{\partial #1}{\partial #2}}

\newcommand\Esp{\mathbb{E}}
\newcommand\Var{\mathbb{V}ar}
\newcommand\Gaus{\mathcal{N}}
\newcommand\trans[1]{{#1}^\intercal}

\RequirePackage[makeroom]{cancel}

\author{Brice Ozenne, Patrick M. Fisher, Esben Budtz-J\o{}rgensen}
\date{}
\title{Small sample corrections for Wald tests in Latent Variable Models}
\hypersetup{
 colorlinks=true,
 citecolor=[rgb]{0,0.5,0},
 urlcolor=[rgb]{0,0,0.5},
 linkcolor=[rgb]{0,0,0.5},
 pdfauthor={Brice Ozenne, Patrick M. Fisher, Esben Budtz-J\o{}rgensen},
 pdftitle={Small sample corrections for Wald tests in Latent Variable Models},
 pdfkeywords={},
 pdfsubject={},
 pdfcreator={Emacs 25.2.1 (Org mode 9.0.4)},
 pdflang={English}
 }
\begin{document}


\maketitle

\section*{Abstract}
\label{sec:orgdfeac02}
Latent variable models (LVMs) are commonly used in psychology and
increasingly used for analyzing brain imaging data. Such studies
typically involve a small number of participants (\(n\)<100), where
standard asymptotic results often fail to appropriately control the
type 1 error. This paper presents two corrections improving the
control of the type 1 error of Wald tests in LVMs estimated using
maximum likelihood (ML). First, we derive a correction for the bias of
the ML estimator of the variance parameters. This enables us to
estimate corrected standard errors for model parameters and corrected
Wald statistics. Second, we use a Student's \(t\)-distribution instead
of a Gaussian distribution to account for the variability of the
variance estimator. The degrees of freedom of the Student's
\(t\)-distributions are estimated using a Satterthwaite
approximation. A simulation study based on data from two published
brain imaging studies demonstrates that combining these two
corrections provides superior control of the type 1 error rate
compared to the uncorrected Wald test, despite being conservative for
some parameters. The proposed methods are implemented in the R package
lavaSearch2 available at
\url{https://cran.r-project.org/web/packages/lavaSearch2}.

\bigskip

\textbf{keywords}: latent variable models, maximum likelihood, repeated
measurements, small sample inference, Wald test.

\section{Introduction}
\label{sec:orgb9de0e6}

\begin{figure}[!b]
\centering
\includegraphics[width=0.6\textwidth]{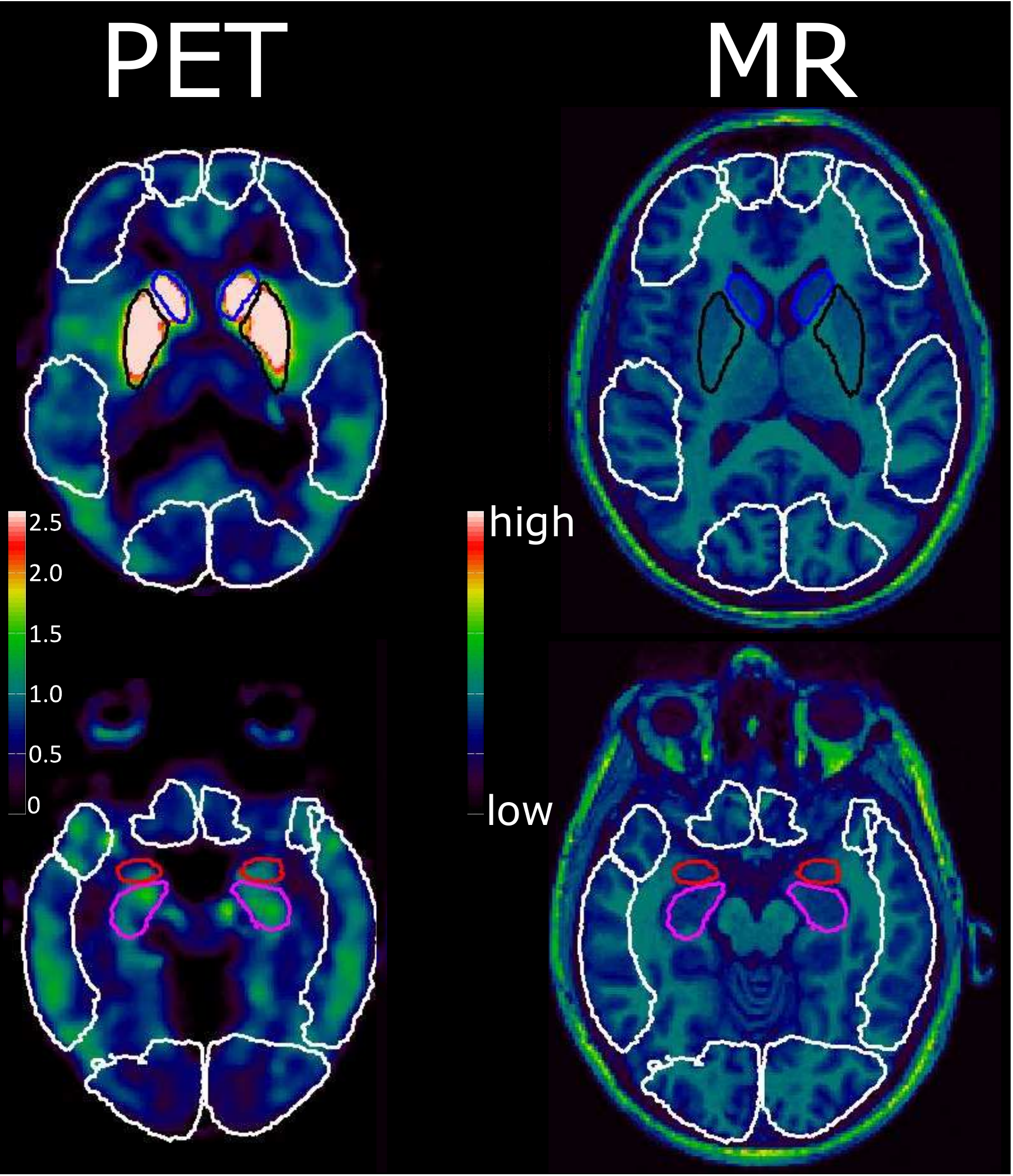}
\caption{\label{fig:PET}
Single-subject serotonin 4 receptor binding and five regions of interest (ROI): amygdala (red areas), caudate (blue), hippocampus (purple), neocortex (white) and putamen (black). Images at left show ROIs overlaid on a brain image depicting voxel-level serotonin 4 receptor binding (color scale units, serotonin 4 receptor binding). Images at right show the same regions overlaid on a high-resolution structural MR image, for spatial orientation.}
\end{figure}

Understanding brain mechanisms is essential to improve prevention and
treatment of brain disorders. For instance, it has been hypothesized
that the serotonin system is a key factor in major depressive
disorders (MDD) and most antidepressants attempt to act on this
system. Unfortunately, less than half of the patients respond to
first-line antidepressant treatment. A deeper understanding of the
serotonin brain system is therefore needed. While it is not yet
possible to non-invasively measure the extracellular level of
serotonin in the brain, medical imaging allows one to simultaneously
visualize the brain structure (using Magnetic Resonance Imaging - MRI)
and quantify the binding potential of certain serotonin receptors
(using Positron Emission Tomography - PET), e.g. see \autoref{fig:PET}.

\bigskip

Recently, Latent variable models (LVMs) have been used to identify the
  brain serotonin level from the binding potentials measured in
  several brain regions and relate this level to patient group, test
  performance, or genotype status (\citet{fisher2015bdnf},
  \citet{fisher2017bdnf}, \citet{stenbaek2017brain}, \citet{deen2017low},
  \citet{perfalk2017testosterone}, \citet{da2018men}). These models were
  estimated by maximum likelihood (ML) and statistical inference was
  most often performed based on the asymptotic distribution of Wald
  statistics. However, the sample size in these studies is rather
  limited (respectively 68, 144, 24, 34, 41, 43), especially in light
  of the number of parameters required to obtain a satisfying model
  fit (respectively 29, 29, 48, 31, 37, 40). The application of
  asymptotic results is thus questionable, e.g., one may be worried
  that the type 1 error is not at its nominal level. This has been
  shown using simulation studies for the global fit tests (i.e.,
  likelihood ratio test, \cite{herzog2007model}), for which corrections
  have been proposed
  \citep{satorra1994corrections,bentler1999structural,wu2016scaled,jiang2017four,maydeu2017maximum}.
  To our knowledge, the small sample properties of the Wald test in
  LVMs has not been carefully studied and software packages for LVMs,
  such as the R package lavaan \citep{lavaan} or Mplus \citep{Mplus},
  implement several small sample corrections for the global fit tests,
  but no corrections for the Wald tests.

\bigskip 

Current solutions for small sample inference include profile
likelihood \citep{pek2015profile}, the use of resampling procedures:
bootstrap, permutation, jackknife, or the use of Bayesian techniques
such as Monte Carlo Markov Chains (MCMC). The main drawback of these
methods is that they are computationally intensive. In addition, each
method has specific pros and cons. For instance, bootstrap and
jackknife may not appropriately control the type 1 error rate because
they rely on asymptotic results, e.g., \citet{parr1983note} and
\citet{carpenter2000bootstrap}. Although permutation procedures
appropriately control the type 1 error rate, they can test only very
specific combinations of the model parameters. \citet{mcneish2016using}
have shown that MCMC is highly sensitive to the specification of the
prior distributions of the parameters in small samples and may,
therefore, not be straightforward to use. In this article we focus on
LVMs estimated by ML and propose an analytical approach to approximate
the distribution of the Wald statistics. This approach does not
require any user input nor any additional model fit. It modifies the
usual asymptotic distribution of the Wald statistic in two ways,
similar to the correction proposed by \citet{kenward1997small} for mixed
models: (i) correcting the bias of the ML-estimator for variance
parameters and (ii) modeling the distribution of the Wald statistics
using Student's \(t\)-distributions instead of Gaussian distributions.

\bigskip

The remainder of the article is structured as follows: we formally
introduce LVMs and discuss the validity of the traditional testing
procedure in section \ref{notationLVM}. In section \ref{applicationLVM}, we
illustrate the use of LVM and the inflated type 1 error rate of the
traditional testing procedure in three applications. Our small sample
correction is presented in sections \ref{biasCorrection} and
\ref{dfEstimation}. They, respectively, extend (i) and (ii) when testing a
single hypothesis in LVMs. Extension to multiple hypotheses and robust
standard errors are discussed in section \ref{Extensions}. The control of
the type 1 error rate after correction is assessed in section
\ref{simulationStudy} using simulations studies inspired from the three
applications. These are re-analyzed with the proposed correction in
section \ref{applicationSSC}. We end the article with a discussion. The
proposed correction is implemented in an R package called lavaSearch2,
available on CRAN
(\url{https://cran.r-project.org/web/packages/lavaSearch2}). An overview of
the functionnalities and code examples can be found in the vignette of
the package. The code used for the simulation studies and for the
illustrations is available at
\url{https://github.com/bozenne/Article-lvm-small-sample-inference}.

\section{Inference in linear LVMs}
\label{notationLVM}
We consider a sample of \(n\) independent and identically distributed
(iid) replicates \((\VY_i,\VX_i)_{i\in\{1,\ldots,n\}}\) generated
by \(m\) endogenous random variables \(\VY=(\Y_1, \dots, \Y_m)\) and
\(l\) exogenous random variables \(\VX=(\X_1, \dots, \X_l)\). A LVM is
defined by a measurement model linking the endogenous variables to a
set of latent variables \(\Veta\) and to the exogenous variables:
\begin{align}
\VY_i &= \boldsymbol{\nu} + \Veta_i \Lambda + \VX_i K + \Vvarepsilon_i \text{, where } \Vvarepsilon_i \sim \Gaus\left(0,\Sigma_\varepsilon\right)
\label{eq:measurement}
\end{align}
and by a structural model relating the latent variables to the
exogenous variables:
\begin{align}
\Veta_i &= \boldsymbol{\alpha} + \Veta_i B + \VX_i \Gamma + \boldsymbol{\zeta}_i \text{, where } \boldsymbol{\zeta}_i \sim \Gaus\left(0,\Sigma_\zeta\right)
\label{eq:structural}
\end{align}
where \(B\) is a matrix with 0 on its diagonal and such that \(1-B\)
is invertible. We denote by \(p\) the number of parameters, by
\(\Vparam\) the vector containing the model parameters (we use the
bold notation to denote row vectors), and by \(\set_{\Vparam}\) the
set of model parameters. The conditional distribution of \(\VY\) given
\(\VX\) follows from equations \eqref{eq:measurement} and
\eqref{eq:structural}:
\begin{align}
& \VY_i|\VX_i \sim \Gaus\left(\Vmu(\Vparam,\VX_i),\Omega(\Vparam)\right) \notag \\
\text{where } & \Vmu(\Vparam,\VX_i) = \boldsymbol{\nu} + \boldsymbol{\alpha} (1-B)^{-1} \Lambda 
+ \VX_i \Gamma (1-B)^{-1} \Lambda  + \VX_i K \notag \\
\text{and } & \Omega(\Vparam) = \trans{\Lambda} (1-B)^{-\intercal} \Sigma_\zeta (1-B)^{-1} \Lambda + \Sigma_\varepsilon \label{eq:LVM_Omega}
\end{align}
The parameters can either be involved in the conditional mean, both in
the conditional mean and variance, or only in the conditional
variance. Parameters of the first type, i.e., parameters
\(\boldsymbol{\nu}\), \(\boldsymbol{\alpha}\), \(K\) and \(\Gamma\),
will be called mean parameters and denoted \(\Vparam_\mu\). Parameters
satisfying the latter type, i.e., parameters in \(\Sigma_\varepsilon\)
and \(\Sigma_\zeta\), will be called variance parameters and denoted
\(\Vparam_\Sigma\). Estimation can be carried out using ML, see
\citet{holst2013linear} for more details. ML is known to give
asymptotically unbiased, efficient and normally distributed
estimates. For a given parameter \(\param \in \set_{\Vparam}\), we can
use a Wald statistic:
\begin{equation}
 t_{\param} = \frac{\paramHat}{\sigma_{\paramHat}} \label{eq:Wald}
\end{equation}
to assess whether \(\param = 0\). Under the null hypothesis,
\(t_{\param}\) is asymptotically normally distributed with mean 0 and
variance 1. Here \(\paramHat\) denotes the value of \(\param\)
estimated using ML and \(\sigma_{\paramHat}\) is the standard
deviation of the estimator (the variance-covariance matrix of the
estimator of \(\Vparam\) will be denoted \(\Sigma_{\VparamHat}\)). In
most applications \(\sigma_{\paramHat}\) is not known but we can plug
the ML estimate of the standard error, \(\hat{\sigma}_{\paramHat}\),
in equation \eqref{eq:Wald} to obtain a tractable test statistic:
\begin{equation}
 \hat{t}_{\param} = \frac{\paramHat}{\hat{\sigma}_{\paramHat}} \label{eq:Wald2}
\end{equation}
This has two consequences in finite samples: (i) the variance of
 \(\hat{t}_{\param}\) will typically be greater than 1 and (ii)
 \(\hat{t}_{\param}\) may not be normally distributed due to the
 variability of \(\hat{\sigma}_{\paramHat}\). Indeed, if
 \(\hat{\sigma}^2_{\paramHat}\) follows a \(\chi^2\) distribution
 and is independent of \(\paramHat\), then \(\hat{t}_{\param}\)
 follows a Student's \(t\)-distribution (up to a multiplicative
 factor). Regarding (i), using a first order Taylor expansion and
 taking the expectation, we can express the first order bias of
 \(\hat{\sigma}_{\paramHat}\):
\begin{align}
\Esp\left[\hat{\sigma}_{\paramHat} -\sigma_{\paramHat}\right] = 
 \Esp\left[\VparamHat-\Vparam\right]\trans{\dpartial{\sigma_{\paramHat}}{\Vparam}}
+ o_p(n^{-\half}) \label{eq:biasVcov}
\end{align}
In correctly specified models, \(\sigma_{\paramHat}\) can be
consistently estimated using the appropriate element in the inverse of
the expected information matrix (denoted \(\Information(\Vparam)\)). As
shown in supplementary material \ref{SM-SM:Information}, the expected
information relative to the parameters \(\param\) and \(\param'\) in a
LVM can be expressed as:
\begin{align}
\Information(\param,\param') =& \frac{n}{2} tr\left(\Omega(\Vparam)^{-1} \dpartial{\Omega(\Vparam)}{\param} \Omega(\Vparam)^{-1} \dpartial{\Omega(\Vparam)}{\param'}\right) 
 + \sum_{i=1}^n \dpartial{\Vmu(\Vparam,\VX_i)}{\param} \Omega(\Vparam)^{-1} \trans{\dpartial{\Vmu(\Vparam,\VX_i)}{\param'}} \label{eq:Information}
\end{align}
Since \(\sigma_{\paramHat}\) is an element of the inverse of
\(\Information(\Vparam)\), it depends on the variance parameters via
\(\Omega(\Vparam)\) so \(\dpartial{\sigma_{\paramHat}}{\Vparam}\) is
typically non-zero. In small samples, the ML estimator of the variance
parameters is in general biased \citep{harville1977maximum}; it follows
that the first term of equation \eqref{eq:biasVcov} is non-zero and
\(\hat{\sigma}_{\paramHat}\) is biased in finite samples. We expect
that the ML estimator of the variance parameters will be biased
downward and \(\sigma_{\paramHat}\) will increase with the variance
parameters, so \(\hat{\sigma}_{\paramHat}\) will be biased downward.

\section{Applications to real data}
\label{applicationLVM}
\begin{figure}[!b]
    \centering
    \begin{subfigure}[!h]{0.45\textwidth}
        \centering
        \includegraphics[width=\linewidth]{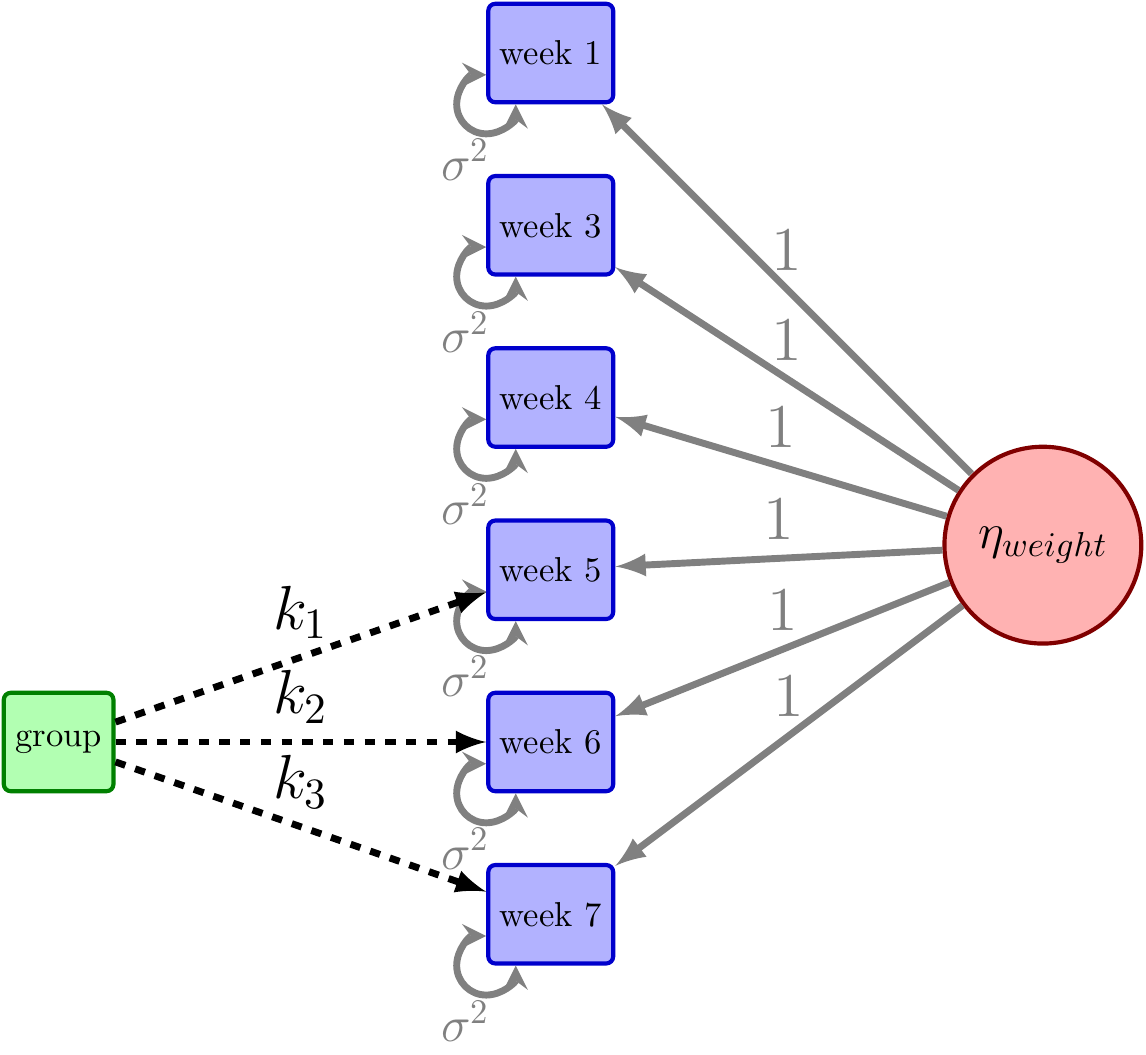}
        \caption{: Growth of guinea pigs} \label{fig:graphPig}
        \includegraphics[width=\linewidth]{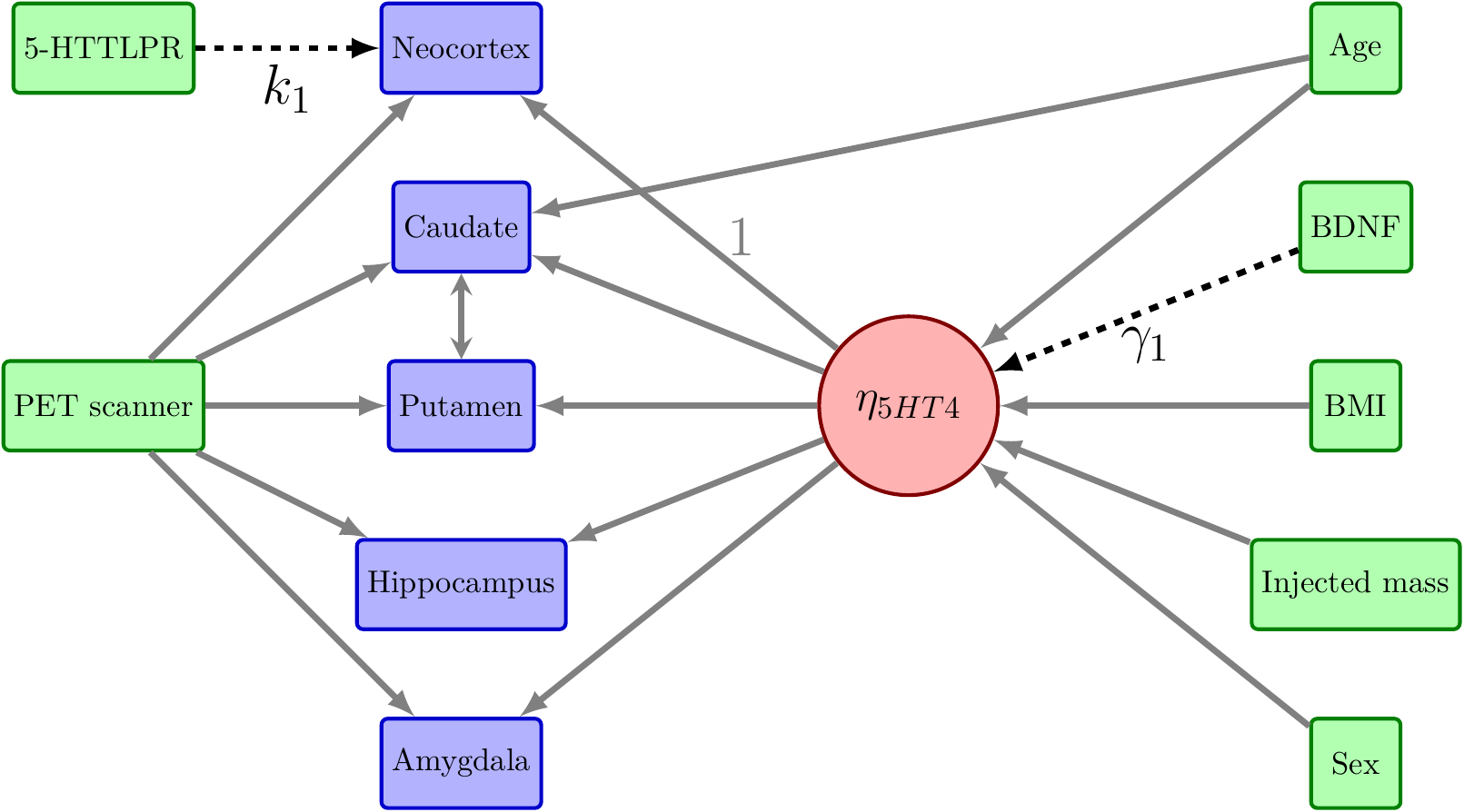} 
        \caption{: serotonin 4 receptor and genetric polymorphisms \citep{fisher2015bdnf}} \label{fig:graphFisher}
    \end{subfigure}
    \begin{subfigure}[!h]{0.45\textwidth}
    \centering
        \includegraphics[width=\linewidth]{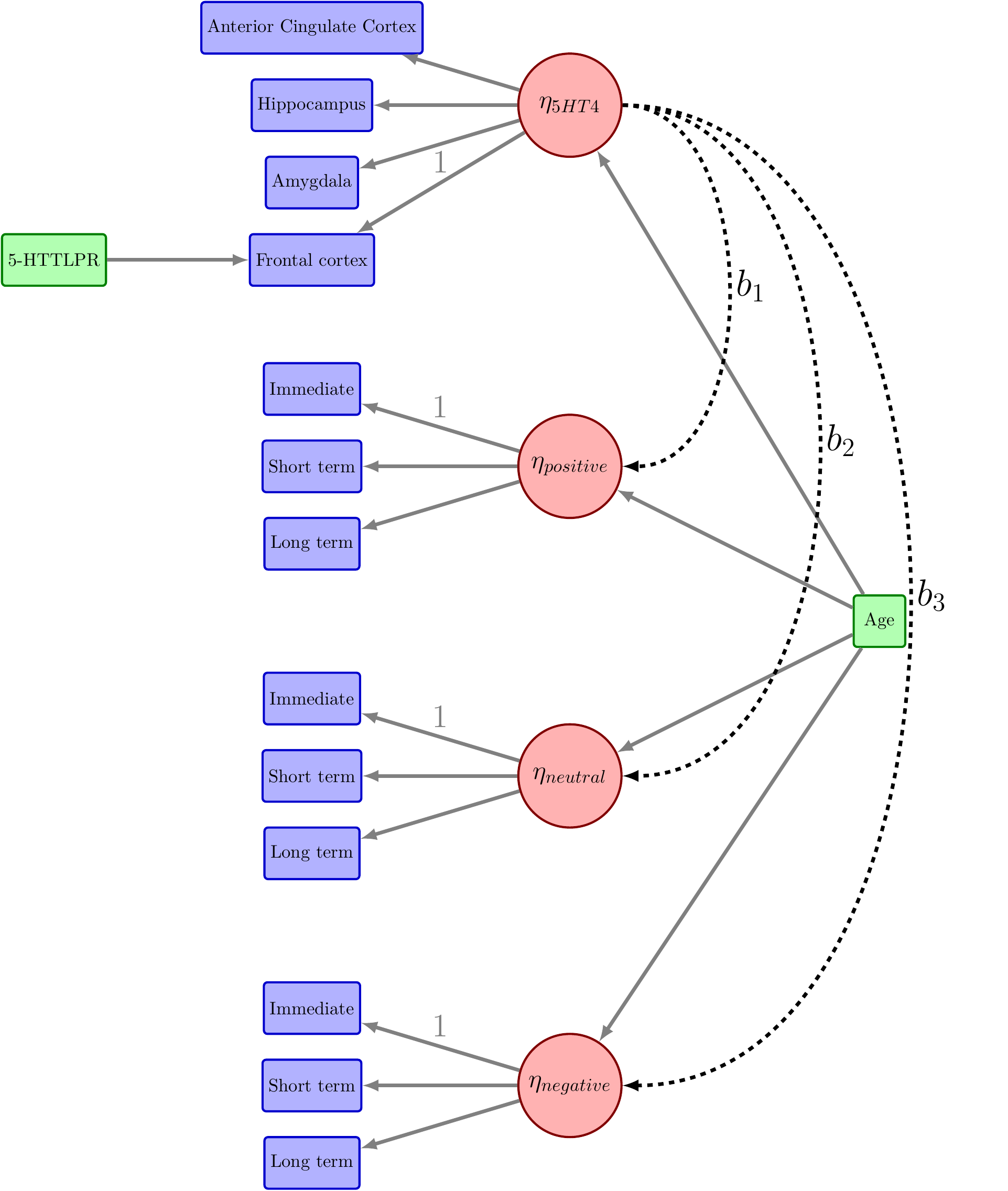} 
        \caption{: serotonin 4 receptor and verbal memory recall \citep{stenbaek2017brain}} \label{fig:graphDea}
    \end{subfigure}
    \caption{Path diagram of the LVM associated with each application: 
     the observed variables are displayed in blue (outcomes) and green (covariates) while the latent variable is displayed in red.
     The union of the full and dotted arrows represent the relationships between variables modeled by the LVM.
     The black dotted arrows denote the parameters of interest and symbols above the full arrows represent constraints, 
     e.g., in \subref{fig:graphPig} the residual variances are constrained to have the same value and the loadings to be 1.} 
    \label{fig:graphApplication}
\end{figure}

We consider three applications involving simple to complex LVMs (see
\autoref{fig:graphApplication} for the corresponding path diagrams). The
first investigates the correspondance between mixed models and LVMs
and the next two were chosen from the studies on the serotonin system
mentioned in the introduction. The latter two applications are
representative of scientific questions encountered in that field: can
we correlate a genetic variable to indirect measurements of the brain
serotonin system? Can we study the relationship between indirect
measurements of the brain serotonin system and indirect measurements
of the cognitive ability of a subject?

\renewcommand{\thesubfigure}{Application \Alph{subfigure}}

\bigskip

\textbf{Growth of guinea pigs (Application A)}: this application originates
from a practical class on mixed models
(\url{http://publicifsv.sund.ku.dk/\~jufo/RepeatedMeasures2017.html}) where
the students are asked to monitor the growth of two groups of guinea
pigs over seven weeks using a random intercept model. One group of
pigs received vitamin E at the beginning of week 5, while the other
group serves as a control (see supplementary material
\ref{SM-SM:figPig} for a graphical display of the data) so the
treatment effect is only modeled from week 5 to 7 using a different
parameter for each week. The dataset used is small: each group
contains only five animals. The random intercept model is equivalent
to a LVM where the loadings are set to 1
(i.e. \(\Lambda=(\lambda_1,\ldots,\lambda_6)=(1,\ldots,1)\)) and the
residual variances are assumed constant over time:
\begin{align*}
Y_{it} &= \left\{ \begin{array}{cc}
\nu_t + \eta_i + \varepsilon_{it} \text{, if } t<5 \\
\nu_t + \eta_i + \text{group}_i k_{t-4} + \varepsilon_{it} \text{, if }  t \geq 5 
\end{array}\right.
 \text{ where } \varepsilon_{it} \sim \Gaus\left(0,\sigma^2\right) \text{ and } \eta_{i} \sim \Gaus\left(0,\tau\right)
\end{align*}
The corresponding path diagram is shown in the upper left panel of
\autoref{fig:graphApplication}.

The interest lies in assessing whether vitamin E affects the growth of
the pigs. This can be performed by testing whether all treatment
parameters are 0 (i.e. \(k_1=k_2=k_3=0\)) using a Wald test. In a
mixed model, we obtained a Wald statistic of 4.25 and a p-value of
0.0102 while the Wald statistic of the LVM was 5.02 and the
corresponding p-value was 0.00176. The large discrepancy is due to the
fact that, in mixed models, restricted maximum likelihood (REML) is
preferred over ML when estimating the model because it corrects the
bias of the ML estimator of the variance
parameters. \citet{kenward1997small} proposed an additional correction
by modeling the distribution of the Wald statistic using a Student's
\(t\)-distribution and estimating its degrees of freedom by the
method of moments. This has become the gold standard procedure in
mixed models and is what has been used here. This correction is not
available for LVM, leading, in this example, to an inflated type 1
error rate: 0.102 instead of 0.05, according to simulation studies.

\bigskip

\textbf{Serotonin 4 receptor binding and genetic polymorphisms (Application
B)}: \citet{fisher2015bdnf} were interested in whether two genetic
variants, i.e., BDNF val66met and 5-HTTLPR polymorphisms, were
associated with brain serotonin levels. The authors collected genetic
data and PET brain imaging data, using two scanners, from 68 healthy
humans. The final dataset contained 73 observations because five
participants were scanned twice. The serotonin 4 receptor binding, a
proxy for brain serotonin levels, was computed in five brain regions
(amygdala, caudate, hippocampus, neocortex and putamen). Preliminary
regression analyses suggested an association between the BDNF val66met
genotype and serotonin 4 receptor levels in all brain regions, whereas
the effect of the 5-HTTLPR polymorphisms was found to be specific to
only the neocortex region. Data were subsequently analyzed in a LVM
where the regional serotonin 4 binding measurements were linked to a
single latent variable, representing an unobservable brain serotonin
level. This latent variable was affected by BDNF val66met genotype
status to model a global effect (\(\Gamma\) parameter), whereas the
5-HTTPLR effect was directly modeled on the neocortex region (\(K\)
parameter). Covariates assumed to affect the serotonin level were
related to the latent variable. To remove systematic differences in
serotonin measurements, a direct effect of the scanner on each brain
region was modeled. The path diagram of the LVM is displayed in lower
left panel of \autoref{fig:graphApplication}. The LVM had 29 parameters.

Inference was performed using cluster robust standard errors, i.e.,
the score was computed at the participant level when using the
sandwich estimator. The effects of the BDNF val66met and 5-HTTLPR
polymorphisms were estimated to be, respectively,
\(\hat{\gamma}_2=0.074\) (\(\text{p-value}=0.005\)) and
\(\hat{k}_1=-0.073\) (\(\text{p-value}=7.2 \, 10^{-6}\)). In a LVM
estimated under the constraint of no genetic effects as a generative
model, a simulation study showed that the actual type 1 error rate was
0.074 for the effect of the BDNF val66met and 0.061 for the effect of
the 5-HTTLPR polymorphisms.

\bigskip

\textbf{Serotonin 4 receptor binding and verbal memory recall (Application
C)}: \citet{stenbaek2017brain} investigated the relationship between
episodic memory and serotonin 4 receptor binding. They collected data
from 24 healthy volunteers who underwent a PET scan and a verbal
affective memory test (VAMT). PET and memory measures were acquired
proximal to each other but not at the same time. In the VAMT test,
subjects are asked to remember words immediately after having learned
them (immediate memory), five minutes after (short term memory), and
30 minutes after (long term memory). Words were divided into three
categories: positive, negative, and neutral valence words. The
serotonin 4 receptor binding was measured in four brain regions known
to be involved in affective processing and memory (amygdala, anterior
cingulate cortex, frontal cortex and hippocampus). Four latent
variables were constructed: three that combine the immediate, short
term, and long term memory for each type of word and one that combines
the serotonin 4 receptor binding across the brain regions
(respectively, \(\eta_{positive}\), \(\eta_{neural}\),
\(\eta_{negative}\), and \(\eta_{5HT4}\)). Associations between the
latent variables were adjusted for age. The path diagram of the LVM is
shown in \autoref{fig:graphApplication} (right panel). The resulting LVM
had 48 parameters.

With this LVM, they found an effect of the binding construct
(\(\eta_{5HT4}\)) on the memory constructs
(\(\eta_{positive}\), \(\eta_{neutral}\),
\(\eta_{negative}\)) with, respectively, \(\hat{b}_1=-7.3\)
(\(\text{p-value}= 0.0005\)), \(\hat{b}_2=-6.7\) (\(\text{p-value}=
0.004\)), and \(\hat{b}_3=-3.7\) (\(\text{p-value}=0.07\)). Using a
LVM estimated under the constraint of no relation between memory and
serotonin 4 receptor binding, a simulation study found that the type 1
error was 0.063 for \(b_1\), 0.085 for \(b_2\), and 0.084 for \(b_3\).

\section{Bias correction for the ML-estimator of the variance parameters}
\label{biasCorrection}
We now develop a method to correct the small sample bias of the estimated
variance parameters, \(\VparamHat_\Sigma\). Denoting the observed
residuals by \(\Vxi_i(\VparamHat)=\VY_i-\Vmu(\VparamHat,\VX_i)\), it
is well known that in a standard linear model the variance of the
observed residuals underestimates the (true) conditional variance of
\(Y\). We show in supplementary material \ref{SM-SM:varResiduals} that
this result generalizes to LVMs. Indeed, given that
\(\Esp\left[\trans{\Vxi_i(\Vparam)}\Vxi_i(\Vparam)\right]=\Omega(\Vparam)\)
and \(\Information(\Vparam)^{-1} = \Sigma_{\VparamHat}\), the
variance of the observed residuals can be decomposed into:
\begin{align}
\Esp\left[\trans{\Vxi_i(\VparamHat)}\Vxi_i(\VparamHat)\right] &= \Omega(\Vparam) - \Psi_i + o_p(n^{-1})  
\quad \text{ where }
\Psi_i = \trans{\dpartial{\Vmu(\Vparam,\VX_i)}{\Vparam}} \Sigma_{\VparamHat} \dpartial{\Vmu(\Vparam,\VX_i)}{\Vparam} \label{eq:biasOmega}
\end{align}
Since the first order bias, \(\Psi_i\), is positive definite, the
variance of the observed residuals is a downward biased estimate of
\(\Omega(\Vparam)\). This result is similar to the one found by \citet{goran2001note}
for Generalized Estimating Equation models. Denoting \(\Psi =
\frac{1}{n} \sum_{i=1}^n \Psi_i\), we obtain by averaging over the
samples:
\begin{align*}
\Esp\left[ \frac{1}{n} \sum_{i=1}^n \trans{\Vxi_i(\VparamHat)}\Vxi_i(\VparamHat) \right] = \Omega(\Vparam) - \Psi + o_p(n^{-1})
\end{align*}

\bigskip

\uline{Example (standard linear model):} consider the generative mechanism
\(Y_i = \VX_i \beta + \varepsilon_i\) with \(\varepsilon_i \sim
\Gaus\left(0,\sigma^2\right)\), where \(Y\) is a univariate
endogenous variable, \(\VX\) contains \(p\) exogenous variables and
\(\varepsilon_i\) are independent and identically normally
distributed. As shown in supplementary material
\ref{SM-SM:LM-correctBias}, formula \eqref{eq:biasOmega} gives that \(\Psi_i = \sigma^2 \VX_i \left(\trans{\VX} \VX\right)^{-1}\trans{\VX}_i\) and \(\Psi = \frac{p}{n}\sigma^2\). Note that the first order bias
of the ML estimator can be removed by considering the estimator
\(\left(\hat{\sigma}^c\right)^2 = \hat{\sigma}^2 + \hat{\Psi} =
(1+\frac{p}{n})\hat{\sigma}^2\).

\bigskip

This example motivates the use of \(\Psi\) to correct the small sample
bias of the ML estimator. We only attempt to correct the bias of the
variance parameters, \(\Vparam_\Sigma\), because simulation studies
show a much smaller bias for the other parameters (e.g., see
supplementary material \ref{SM-SM:tableBiasML}). To do so, we assume
that \(\Omega(\VparamHat)\) and \(\Esp\left[ \frac{1}{n} \sum_{i=1}^n
\trans{\Vxi_i(\VparamHat)}\Vxi_i(\VparamHat) \right]\) have the same
first order bias. Then, given \(\Psi\), we can defined a corrected
ML-estimator of \(\Omega\): \(\hat{\Omega}^c = \Omega(\VparamHat) +
\Psi\). From equation \eqref{eq:LVM_Omega} and considering
\((\Lambda,B)\) fixed, we get that \(\Omega(\Vparam)\) is linearly
related to the variance parameters, so we can find a matrix \(Z\)
(depending only on \(\Lambda\) and \(B\)) such that
\(vec(\hat{\Omega}^c) = Z \, \Vparam_\Sigma + \mathbf{r}\), where
\(\mathbf{r}\) is a possible residual error and \(vec\) is the column
stacking operator transforming \(\hat{\Omega}^c\) into a vector. Given
\(Z\), we obtain a new estimate of \(\Vparam_\Sigma\) by solving the
least squares problem. This leads to the following iterative procedure
to estimate \(\Psi\) and get bias-corrected estimates of
\(\Vparam_\Sigma\) and \(\Information(\Vparam)\):

\RestyleAlgo{boxruled}
{\linespread{2}
\begin{algorithm}[H]
\textbf{Initialize:} 
 \(\hat{\Information}^{(0)} \leftarrow \hat{\Information}(\Vparam)\) \\
\For{k=1 to \(\infty\)}{
(i) Compute for each subject \(i\) the bias
  \(\hat{\Psi}_i^{(k)}\) by plugging
  \(\left(\hat{\Information}^{(k-1)}\right)^{-1}\) and 
  \(\dpartial{\Vmu(\VparamHat,\VX_i)}{\Vparam}\) into
  equation \eqref{eq:biasOmega}. \\ [2mm]
(ii) Compute the average bias \(\hat{\Psi}^{(k)} =
  \frac{1}{n}\sum_{i=1}^n \hat{\Psi}_i^{(k)}\). \\ [2mm]
(iii) Estimate \(\hat{\Omega}^{(k)}\) by \(\Omega(\VparamHat)+\hat{\Psi}^{(k)}\). \\ [2mm]
(iv) Find \(\VparamHat_\Sigma^{(k)}\) satisfying \(\hat{\Omega}^{(k)} = \trans{\hat{\Lambda}} (1-\hat{B})^{-\intercal} \Sigma_\zeta\left(\VparamHat_\Sigma^{(k)}\right) (1-\hat{B})^{-1} \hat{\Lambda} + \Sigma_{\varepsilon} \left(\VparamHat_{\Sigma}^{(k)}\right) \). \\ [2mm]
(v) Compute the corrected derivatives \(\dpartial{\hat{\Omega}^{(k)}}{\Vparam}\) using \(\VparamHat_\Sigma^{(k)}\). \\ [2mm]
(vi) Compute \(\hat{\Information}^{(k)}\) by plugging \(\hat{\Omega}^{(k)}\), 
\(\dpartial{\Vmu(\VparamHat,\VX_i)}{\Vparam}\), 
and \(\dpartial{\hat{\Omega}^{(k)}}{\Vparam}\) into equation \eqref{eq:Information}. \\
}
\caption{}
\end{algorithm}
}

\bigskip

In practice, the algorithm is stopped when the difference between two
consecutive estimates of \(\hat{\Omega}^{(k)}\) is small, e.g., in our
software implementation we require that the Frobenius norm of the
difference is smaller than \(10^{-5}\). We can check that algorithm 1
gives a sequence of \(\hat{\Psi}^{(k)}\) that are semi-definite
positive. Indeed, if at step (i) \(\hat{\Information}^{(k-1)}\) is
definite positive then \(\hat{\Psi}^{(k)}_i\) is semi-definite
positive. This in turn would imply that \(\hat{\Psi}^{(k)}\) is
semi-definite positive, \(\hat{\Omega}^{(k)}\) is definite positive,
and so is \(\hat{\Information}^{(k)}\). We obtain the stated property
by induction. Showing the convergence and monotonicity of
\(\Psi^{(k)}\) is more difficult, so we only consider specific cases:

\bigskip

\uline{Example (standard linear model):} supplementary material
\ref{SM-SM:LM-correctBias} shows that the estimated bias of the
residual variance at iteration \(k\) is \(\hat{\Psi}^{(k)} =
\frac{p}{n} \left(\hat{\sigma}^{(k-1)}\right)^2\). The corresponding corrected residual
variance is \(\left(\hat{\sigma}^{(k)}\right)^2 = \hat{\sigma}^2
\sum_{\kappa=0}^k \frac{p^\kappa}{n^\kappa} \xrightarrow[k\rightarrow \infty]{}
\frac{n}{n-p} \hat{\sigma}^2\). The corrected residual variance tends
toward the usual unbiased estimate of \(\sigma^2\). The convergence is
fast, especially when \(n\) is large, since there is a factor
\(\frac{p}{n}\) between the contribution of the current iteration and
the contribution of the next iteration. The same applies to
\(\hat{\Psi}^{(k)}\). Note that \(\hat{\Psi}^{(k)}\) and
\(\left(\hat{\sigma}^{(k)}\right)^2\) are increasing sequences.

\bigskip

\uline{Example (mean-variance model)}: we consider a LVM where no parameter
appears both in the conditional mean and variance. This corresponds to
common factor models where \(\boldsymbol{\alpha}=0\) and \(\Gamma=0\),
or mixed models where \(B=0\) and \(\Lambda\) is known (e.g. equals 1
for random intercept models as in application A). In both cases,
\(\dpartial{\Omega(\Vparam)}{\param}\) and
\(\dpartial{\mu(\Vparam)}{\param}\) cannot be simultaneously non-0 for
a given parameter, so we obtain from formula \eqref{eq:Information} that
the information matrix is block diagonal. We show in supplementary
material \ref{SM-SM:algo1} that, if the number of mean parameters is
smaller than \(n\), Algorithm 1 converges and \(\hat{\Psi}^{(k)}\)
increases (in the sense of the spectral norm) over iterations.

\bigskip

The corrected estimates of the variance parameters obtained by
Algorithm 1 can be substituted in \(\VparamHat\) to the initial
estimates to obtain \(\VparamHat^c\). As an important side product, we
also obtain a corrected expected information matrix, denoted
\(\hat{\Information}^c\), that can be used to calculate a corrected
Wald statistic.

\bigskip

\uline{Example (standard linear model):} in this model, the variance of the
ML estimator of the regression parameters is \(\Var[\hat{\beta}] =
\frac{\sigma^2}{\trans{X}X}\). Using \(\hat{\Information}^c\) instead
of \(\hat{\Information}\) is equivalent to plug-in
\(\left(\hat{\sigma}^c\right)^2\), an unbiased estimate of
\(\sigma^2\), instead of \(\hat{\sigma}^2\), a downward biased
estimate of \(\sigma^2\). Whereas using Algorithm 1 leads to a
satisfactory estimator for \(\Var[\hat{\beta}]\), the estimator of
\(\Var[\left(\hat{\sigma}^c\right)^2]\) can still be improved. Indeed,
the variance of
\(\left(\hat{\sigma}^c\right)^2=\frac{\sigma^2}{n-p}\sum_{i=1}^n
\frac{\varepsilon_i^2}{\sigma^2}\) equals \(\frac{2 \sigma^4}{n-p}\)
since \(\sum_{i=1}^n \frac{\varepsilon_i^2}{\sigma^2} \sim
\chi^2_{n-p}\). However, the variance estimator obtained after
inverting \(\hat{\Information}^c\) is \(\frac{2
\left(\hat{\sigma}^c\right)^4}{n}\) which is downward biased in finite
samples. We will return to this problem in section \ref{nEstimation}.

\section{Modelling the distribution of the Wald statistics using Student's \(t\)-distributions}
\label{dfEstimation}
In this section we propose a method to account for the uncertainty in
\(\hat{\sigma}^2_{\paramHat}\) when deriving the distribution of
the Wald statistic.

\subsection{Satterthwaite approximation}
\label{sec:orgac2cafc}

In a standard linear model, \(\hat{\sigma}^2_{\paramHat}\) is known
 to be \(\chi^2\) distributed. Indeed
 \(\hat{\sigma}^2_{\paramHat}\) is proportional to the residual
 variance, which can be expressed as the sum of the residuals
 squared. This motivates the use of a Student's \(t\)-distribution
 instead of a Gaussian distribution for the Wald statistic. In
 multivariate models like LVMs, the distribution of
 \(\hat{\sigma}^2_{\paramHat}\) is not generally known. However, it
 can still be approximated using a \(\chi^2\) distribution by finding
 \((\tau,df) \in (\Real,\Real^+)\) such that \(\tau
 \hat{\sigma}^2_{\paramHat} \stackrel{approx.}{\sim} \chi^2(df)\).
 Here \(\tau\) and \(df\) can be identified from the method of
 moments: using that a \(\chi^2\) distribution with \(df\) degrees of
 freedom has expectation \(df\) and variance \(2 df\), we get that
 \(\Esp\left[\tau \hat{\sigma}^2_{\paramHat}\right] = df\) and
 \(\Var\left[\tau \hat{\sigma}^2_{\paramHat}\right] = 2
 df\). Therefore, \(\tau\) and \(df\) should satisfy:
\begin{align}
\tau &= \frac{df}{\Esp\left[\hat{\sigma}^2_{\paramHat}\right]}  
\qquad \qquad \qquad
df = 2 \frac{\Esp\left[\hat{\sigma}^2_{\paramHat}\right]^2}{\Var\left[\hat{\sigma}^2_{\paramHat}\right]}
\label{eq:dfSatterthwaite}
\end{align}
Denoting \(\param^* = \frac{\paramHat}{\sigma_{\paramHat}}\) and
\(\sigma^{*2}_{\paramHat} = \frac{df
 \hat{\sigma}^2_{\paramHat}}{\sigma^2_{\paramHat}}\), one can
re-write the test statistic \(\hat{t}_\param\) as
\(\frac{\param^*}{\sigma^*_{\paramHat}/\sqrt{df}}\).  Under the null
hypothesis, \(\param^* \sim \Gaus\left(0,1\right)\) and
\(\sigma^{*2}_{\paramHat} \sim \chi^2(df)\), \(\hat{t}_\param\)
follows a Student's \(t\)-distribution with \(df\) degrees of
freedom. This approximation is a classical technique in mixed models
and it has been implemented in many software tools, e.g., SAS PROC
MIXED or the R package lmerTest (\cite{kuznetsova2017lmertest}).

\subsection{Application to LVMs}
\label{dfLVMs}
Although we can directly substitute the estimate
\(\hat{\sigma}^2_{\paramHat}\) in equation \eqref{eq:dfSatterthwaite}
for \(\Esp\left[\hat{\sigma}^2_{\paramHat}\right]\), we need an
estimator for \(\Var\left[\hat{\sigma}^2_{\paramHat}\right]\) in order
to estimate \(df\). For a given \(\param \in \set_{\Vparam}\),
\(\hat{\sigma}^2_{\paramHat} = c_j \Information(\VparamHat)^{-1}
\trans{c_j}\) where \(j\) is the index of \(\param\) in
\(\set_{\Vparam}\) and \(c_j\) is a vector with a 1 at the \(j\)-th
position and 0 otherwise. From equation \eqref{eq:Information} we see
that \(\Information(\Vparam)\) depends only on the model parameters
(and on \(\VX_i\), which are fixed values). Using that \(\VparamHat\)
is asymptotically normally distributed with variance
\(\Sigma_{\VparamHat}\), we can apply the multivariate delta method to
obtain an estimator for \(\Var[\hat{\sigma}^2_{\paramHat}]\):
\begin{align*}
n^{1/2} \left(\hat{\sigma}^2_{\paramHat}-\sigma^2_{\paramHat}\right) &\sim \Gaus\left(0, \nabla_{\Vparam} \sigma^2_{\paramHat} \; \Sigma_{\VparamHat} \; \trans{\nabla_{\Vparam} \sigma^2_{\paramHat}}\right)
\end{align*}
Here, \(\nabla_{\Vparam}\) denotes the vector of partial derivatives
relative to each parameter in \(\Vparam\). Therefore \(df\) can be
estimated using the following procedure:
\begin{itemize}
\item for each \(k \in \{1,\ldots,p\}\), compute
\(\dpartial{\Information(\VparamHat)}{\param_k}\), the first
derivative of the information matrix (see supplementary material
\ref{SM-SM:dInformation}).
\item for each \(k \in \{1,\ldots,p\}\), compute
\(\dpartial{\hat{\sigma}^2_{\paramHat_j}}{\param_k}= - c_j
  \Information(\VparamHat)^{-1}
  \dpartial{\Information(\VparamHat)}{\param_k}
  \Information(\VparamHat)^{-1} \trans{c_j}\). Combining all the
partial derivatives into a vector gives \(\nabla_{\Vparam}
  \hat{\sigma}^2_{\paramHat_j}\).
\item estimate the degrees of freedom of \(\hat{\sigma}^2_{\paramHat}\)
as
\(2\frac{\left(\hat{\sigma}^2_{\paramHat_j}\right)^{2}}{\nabla_{\Vparam}
  \hat{\sigma}^2_{\paramHat_j} \; \hat{\Sigma}_{\VparamHat} \;
  \trans{\nabla_{\Vparam} \hat{\sigma}^2_{\paramHat_j}}}\).
\end{itemize}

\bigskip

\uline{Example (standard linear model):} we denote by
\(\sigma^2_{\hat{\beta}}=\frac{\sigma^2}{\trans{X}X}\) the variance of
the estimated regression parameter. The variance of
\(\hat{\sigma}^2_{\hat{\beta}}\) obtained with the delta method is \(2
\frac{\hat{\sigma}^4_{\hat{\beta}}}{n}\). The Satterthwaite
approximation gives
\(df=2\frac{\hat{\sigma}^4_{\hat{\beta}}}{2\hat{\sigma}^4_{\hat{\beta}}/n}=n\)
as the estimate of the degrees of freedom for the Wald statistic
(supplementary material \ref{SM-SM:LM-correctSatterthwaite}). Although
this approximation is better than using a standard normal
distribution, it does not match the true value of \(n-p\) for the
degrees of freedom. The estimator of the residual variance in the
standard linear models is \(\chi^2\) distributed with \(n-p\) degrees
of freedom because the score equation induces \(p\) constraints
between the observed residuals.

\subsection{Effective sample size}
\label{nEstimation}
So far, we have neglected the loss in degrees of freedom caused by the
estimation of the parameters, i.e., using the actual sample size \(n\)
is an upward biased estimator of the number of independent
residuals. This number is used when computing the first term of the
information matrix (equation \eqref{eq:Information}) and, as illustrated
in the previous example, the bias also affects the estimation of the
degrees of freedom of the Wald statistic. We define the effective
sample size as the sum of the dependence of each observed residual on
the corresponding observation:
\begin{align}
\Vn^{c} &= \sum_{i=1}^n \dpartial{\Vxi_i(\VparamHat)}{\VY_i} = \Vn - \sum_{i=1}^n \dpartial{\Vmu(\VparamHat,\VX_i)}{\VY_i} \label{eq:correctedN}
\end{align}
where \(\Vn^c = (n_1^c,\ldots,n_m^c)\) is the vector of effective
sample sizes, with one element per endogenous variable, and \(\Vn =
(n,\ldots,n)\). If the observations would not affect the fit, then
each element of would equal \(n\). However the constraints on the
residuals reduce the variation of \(\Vxi_i(\VparamHat)\) relative to
\(\VY_i\) leading to each element of \(\Vn^c\) being smaller than
\(n\). We see that \(\Vn^{c}\) depends on
\(\dpartial{\Vmu(\VparamHat,\VX_i)}{\VY_i}\), the generalized
leverage, as defined by \citet{wei1998generalized}.

\bigskip

\uline{Example (standard linear model):} we recover the standard result that
the effective sample size is \(n^{c} = n - p\) (supplementary material
\ref{SM-SM:LM-correctDF}). The estimator of the variance of
\(\left(\hat{\sigma}^c\right)^2\) becomes \(\frac{2
\hat{\sigma}^2}{n-p}\) and the degrees of freedom of the Wald
statistic obtained with the Satterthwaite approximation are \(n - p\):
we now have unbiased estimators of the variance of
\(\left(\hat{\sigma}^c\right)^2\) and of the associated degrees of
freedom.

\bigskip

\uline{Example (mean-variance model):} the effective sample size relative to the \(t\)-th endogenous variable can be
expressed as:
\begin{align*}
n - \sum_{i=1}^n \trans{\dpartial{\mu_t(\hat{\Vparam},\VX_i)}{\hat{\Vparam}_\mu}} \left(\sum_{i=1}^n \dpartial{\mu(\hat{\Vparam},\VX_i)}{\hat{\Vparam}_\mu} \Omega(\Vparam)^{-1}\dpartial{\mu(\hat{\Vparam},\VX_i)}{\hat{\Vparam}_\mu}  \right)^{-1} \dpartial{\mu(\hat{\Vparam},\VX_i)}{\hat{\Vparam}_\mu} \Omega(\Vparam)^{-1} c_t
\end{align*}
where \(c_t\) is an \(m\) dimensional vector containing 1 at the
\(t\)-th position and 0 otherwise.

\bigskip

Algorithm 1 can be modified to compute the effective sample size and
obtain corrected degrees of freedom for the Wald statistics
(supplementary material \ref{SM-SM:effectiveSampleSize}).
\section{Extensions}
\label{Extensions}
The Satterthwaite approximation can also be used when considering a
linear combination \(c\) of parameters by substituting \(c \Vparam\)
to \(\param\) and \(c \Sigma_{\VparamHat} \trans{c}\) to
\(\sigma^2_{\paramHat}\) in the expressions presented in section
\ref{dfLVMs}. When simultaneously testing several null hypotheses that
can be defined via a non-singular contrast matrix \(C\) of rank \(Q\),
the Wald statistic becomes:
\begin{align}
\mathcal{F}_{C \Vparam} = \frac{1}{Q} \trans{(C \Vparam)} (C \Sigma_{\VparamHat} \trans{C})^{-1} (C \Vparam) \label{eq:WaldM}
\end{align}
A Satterthwaite approximation can also be derived for this test
 statistic (e.g., see supplementary material
 \ref{SM-SM:multivariateWaldDef}).

\bigskip

Consider a partition of the observations into \(G\) clusters called
\(\mathcal{G}_1,\ldots,\mathcal{G}_G\). Denoting
\(\Score(\Vparam|\VY_i,\VX_i)\) the individual score (see
supplementary material \ref{SM-SM:score} for the mathematical
expression), we can define the score of the \(g\)-th cluster:
\(\Score_{\mathcal{G}_g}(\Vparam) = \sum_{i \in
\mathcal{G}_g}\Score(\Vparam|\VY_i,\VX_i)\). When performing inference
in a misspecified model, \cite{white1982maximum} has shown that the
robust estimator:
\begin{align}
\Sigma^{robust}_{\VparamHat} = \Information(\Vparam)^{-1} \left(\sum_{g=1}^G  \trans{\Score_{\mathcal{G}_g}(\Vparam)} \Score_{\mathcal{G}_g}(\Vparam) \right)  \Information(\Vparam)^{-1}
\label{eq:sandwich}
\end{align}
can consistently estimate the variance of \(\VparamHat\). One
important assumption is that the clusters are iid (assumption A1 in
\cite{white1982maximum}). Robust standard error can then used to obtain
a robust Wald test where the type 1 error is controlled
(asymptotically) even when the normality assumption is violated or
when the covariance structure is partially misspecified (e.g., in
application B, we did not model the correlation between measurements
obtained from the same patients). However, in finite samples, one can
expect that the robust Wald test suffers from the same limitations as
the traditional Wald test. Fortunately, we can apply our bias
correction to the estimator of \(\Sigma^{robust}_{\VparamHat}\) by
plugging the corrected score and information matrix in equation
\eqref{eq:sandwich}. As an approximation, we set the degrees of freedom
of the robust standard errors to be identical to the ones of the
(non-robust) standard errors.

\section{Simulation study}
\label{simulationStudy}
\begin{figure}[!b]
    \centering
    \begin{subfigure}[!h]{0.45\textwidth}
        \centering
        \includegraphics[width=\linewidth]{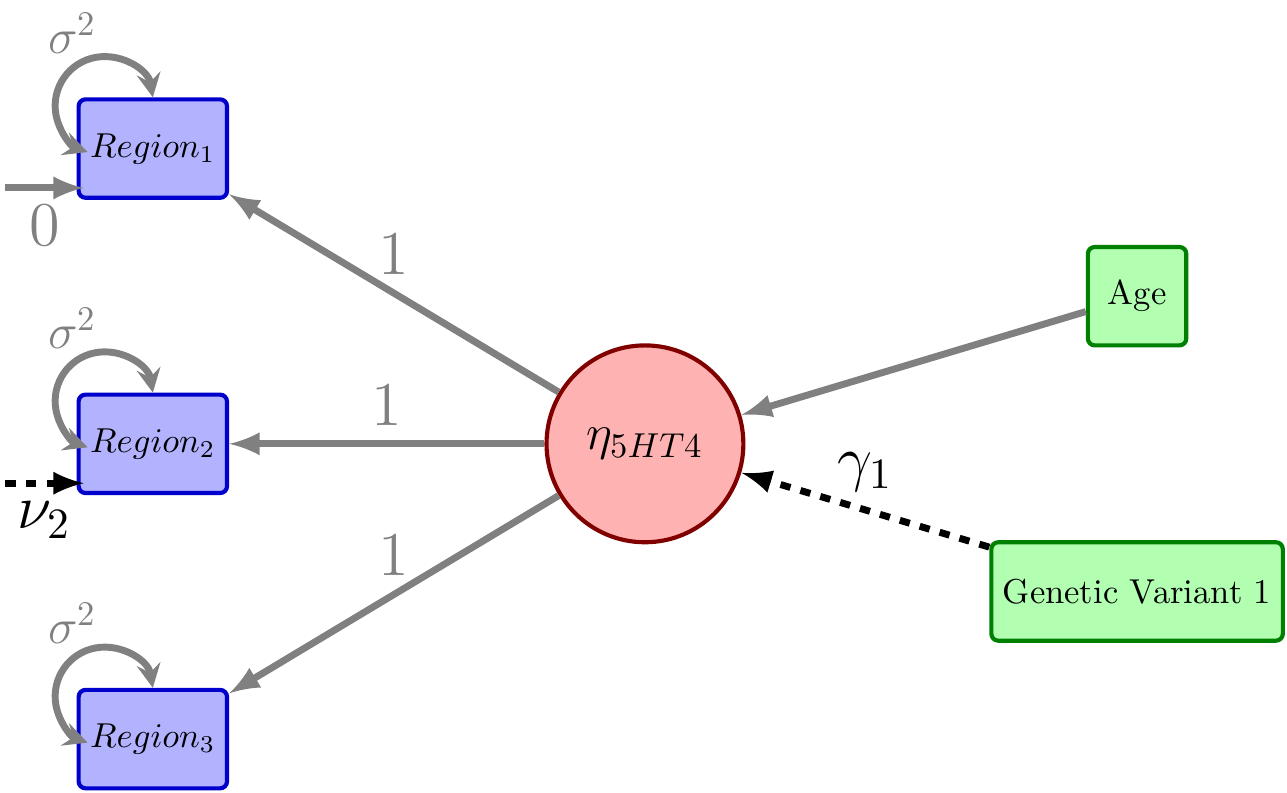} 
        \caption{: mixed model} \label{fig:graphMixedModel}
        \includegraphics[width=\linewidth]{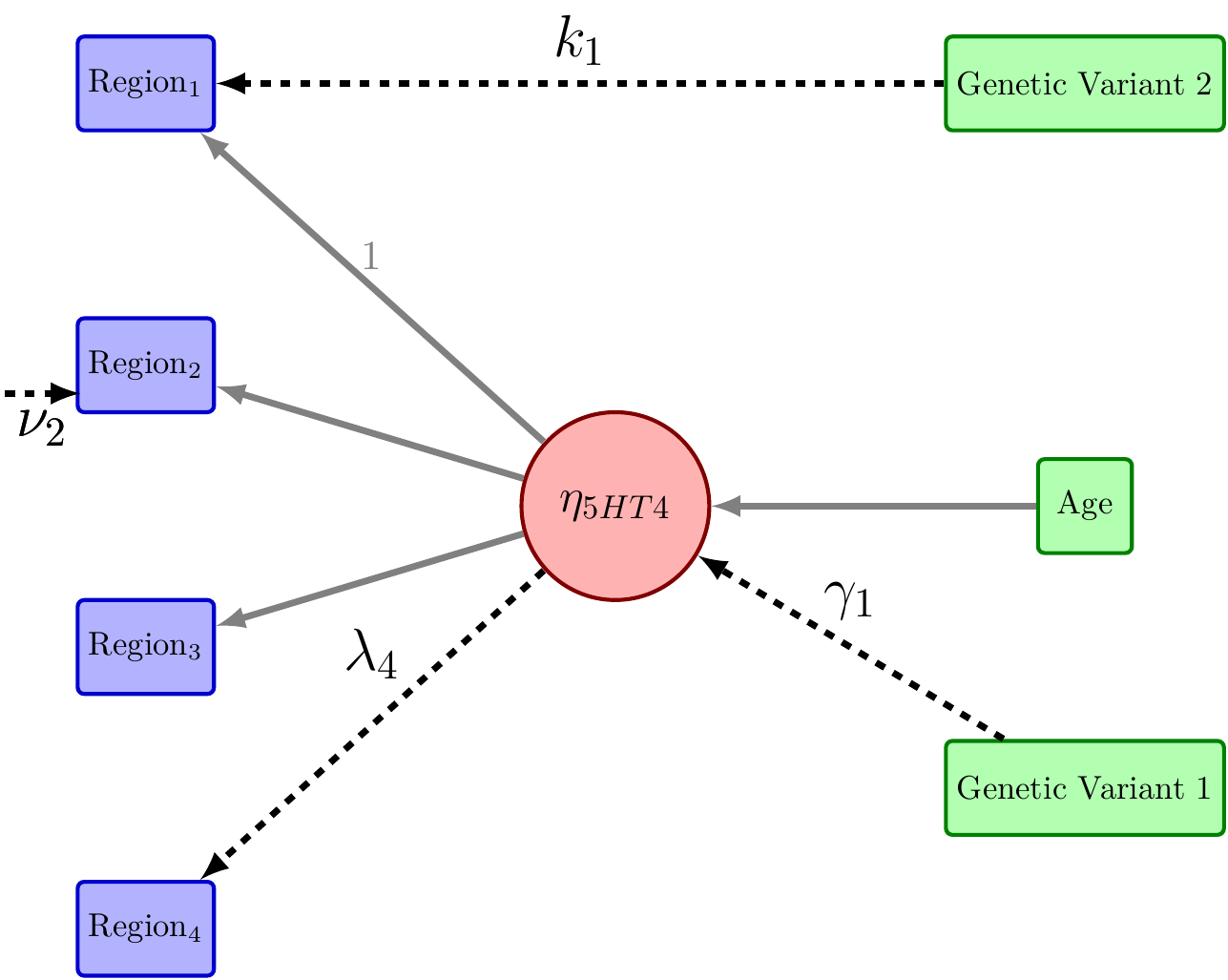} 
        \caption{: single factor model} \label{fig:graphFactorModel}
    \end{subfigure}
    \begin{subfigure}[!ht]{0.45\textwidth}
    \centering
        \includegraphics[width=\linewidth]{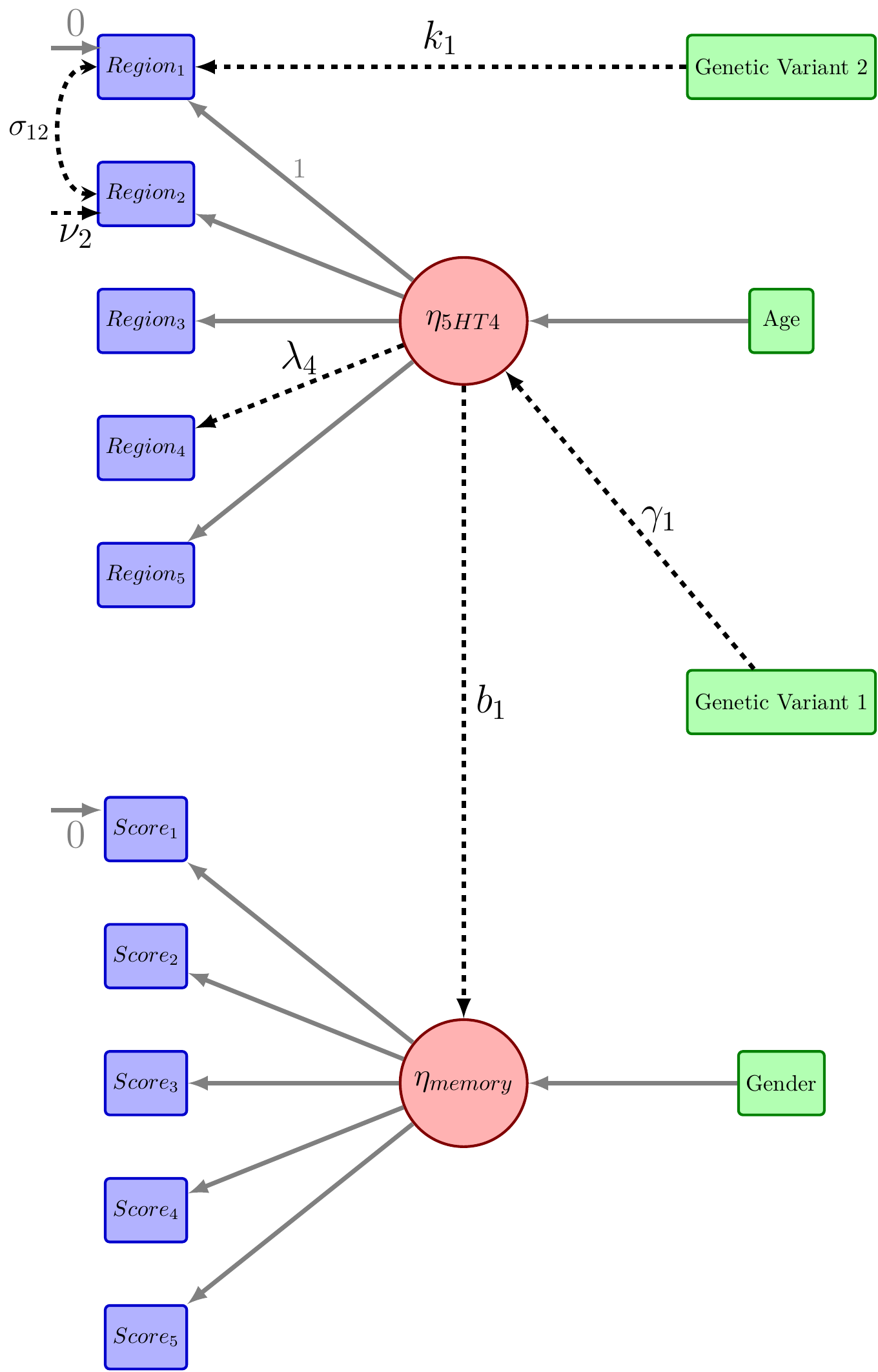} 
        \caption{: two latent variable model} \label{fig:graph2LVM}
    \end{subfigure}
    \caption{Path diagram of the LVM associated with each simulation: 
     endogenous variables are represented in blue,  latent variables in red, and exogenous variables in green.
    The full arrows represent the generative model and symbols above these arrows represent constraints, e.g. in \subref{fig:graphMixedModel} the residual variances of \(Y_1\), \(Y_2\), \(Y_3\) are constrained to have the same value. The values of the parameters in the generative model were 0 for the intercept and 1 for all the other parameters.
    The union of the full and dotted arrows represent the investigator models.
    Wald tests were performed for the parameters represented by a dotted arrow, e.g.
    \(\nu_2\), \(\gamma_1\) in \subref{fig:graphMixedModel}.
    } 
    \label{fig:graphSimulation1Test}
\end{figure}

We performed three simulation studies to investigate the impact of the
proposed corrections on the bias of the estimates and on the control
of the type 1 error. The LVMs used in the simulation studies are
simplified versions of the LVMs used in the real data applications (see
\autoref{fig:graphSimulation1Test} for the corresponding path
diagrams). A simulation study was characterized by (i) a generative
model, i.e., the model defining the distribution used to simulate the
data, (ii) an investigator model, i.e., the model fitted to the
simulated data, (iii) the set of null hypotheses, each testing whether
one of the model parameters equals 0.

\bigskip

For each study, 20 000 samples were generated using the generative
model. Each sample was used to estimate the parameters of the
investigator model. Then, each null hypothesis was tested separately
with a Wald test using 1) the standard procedure (uncorrected
information matrix, Gaussian distribution), 2) the bias correction
(corrected information matrix), 3) the Satterthwaite approximation
(Student's \(t\)-distribution with degrees of freedom estimated
according to section \ref{dfEstimation}), 4) the complete correction,
i.e. bias correction and Satterthwaite approximation. In each case,
the type 1 error was computed as the relative frequency of p-values
lower than 0.05. The small sample bias was assessed for ML estimates
and after application of the bias correction (ML-corrected
estimates). When performing the simulation in small samples, the
estimation algorithms did not always converge. The convergences issues
and how they were handled is detailed in supplementary material
\ref{SM-SM:cvIssues}.

\renewcommand{\thesubfigure}{Study \Alph{subfigure}}

\begin{figure}[!p]
\centering
\includegraphics[width=\textwidth]{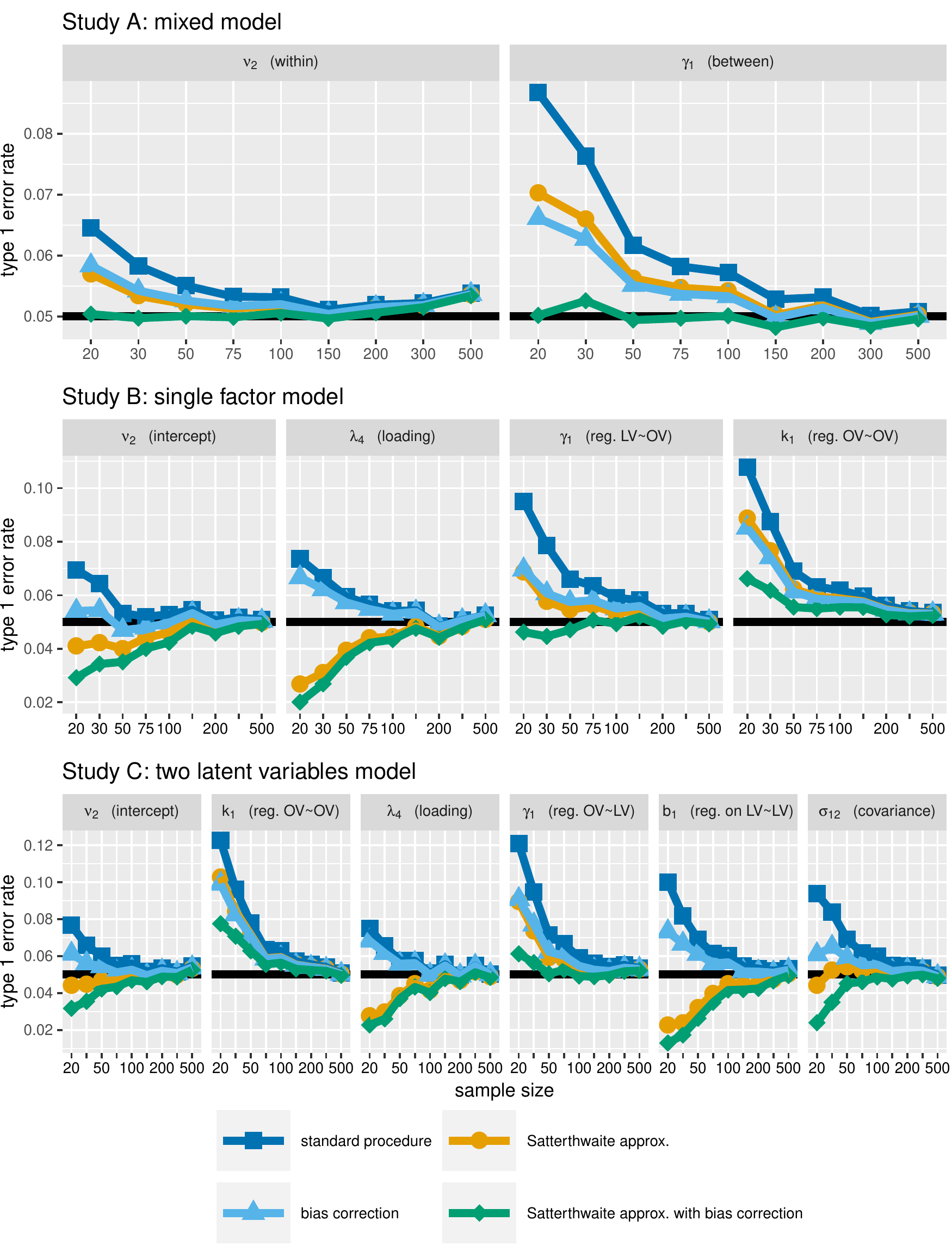}
\caption{\label{fig:type1error-Wald}
Type 1 error rate of the Wald test for study A, B and C at various sample sizes. The type 1 error rates for the parameters that are not shown on the graph (e.g., testing \(\lambda_2=1\)) are similar to the ones shown (e.g., testing \(\lambda_4 = 0\)). At a given sample size, 20000 simulations were performed, meaning that the type 1 error rate is expected to fluctuate between 0.047 and 0.053.}
\end{figure}

\begin{figure}[!p]
\centering
\includegraphics[width=\textwidth]{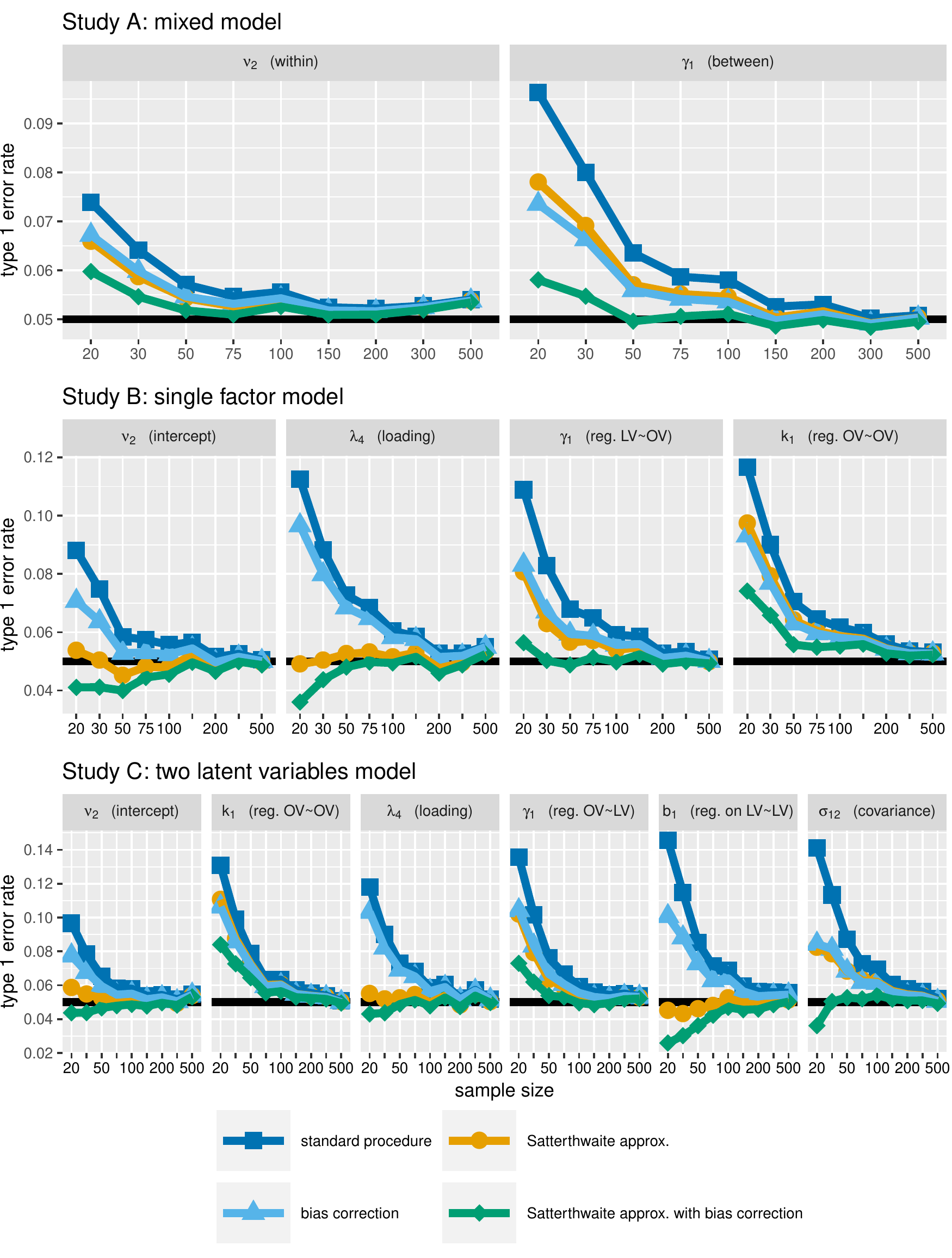}
\caption{\label{fig:type1error-robustWald}
Type 1 error rate of the robust Wald test for study A, B and C at various sample sizes. The type 1 error rates for the parameters that are not shown on the graph (e.g., testing \(\lambda_2=1\)) are similar to the ones shown (e.g., testing \(\lambda_4 = 0\)). At a given sample size, 20000 simulations were performed, meaning that the type 1 error rate is expected to fluctuate between 0.047 and 0.053.}
\end{figure}

\bigskip

\textbf{Wald test in a mixed model (Study A)}: the first LVM is equivalent to
a random intercept model, where the endogenous variable is measured on
three brain regions per subject. The resulting LVM has 7
parameters. The first null hypothesis tests whether the conditional
expectation of the endogenous variable is the same between the first
and second repetition (within subject parameter \(\nu_2\)). The second
null hypothesis tests whether there is an effect of Genetic Variant 1
on all repetitions of the endogenous variable (between subject
parameter \(\gamma_1\)).

The ML estimates of the variance parameters showed a small bias that
was removed by our correction (table \ref{SM-tab:biasMM} in
supplementary material \ref{SM-SM:tableBiasML}). Without correction, a
moderate inflation of the type 1 error rate was observed for the Wald
tests, e.g., \(+0.015\) for \(\nu_2\) and \(+0.037\) for \(\gamma_1\)
when \(n=20\) (first row of \autoref{fig:type1error-Wald}). The bias
correction combined with the Satterthwaite approximation provided a
satisfactory control of the type 1 error rate, e.g., \(0.050\) for
\(\nu_2\) and \(\gamma_1\) when \(n=20\).

\bigskip

\textbf{Robust Wald test in a single factor model (Study B)}: Compared to the
first simulation, the second LVM relaxes the assumption of common
variance and covariance between the measurements, adds another
polymorphism (Genetic Variant 2) and a fourth brain region. The
resulting LVM had 15 parameters. Two additional null hypotheses are
considered: whether the new region is correlated with the first one
(\(\lambda_4\)) and whether there is an effect of Genetic Variant 2 on
the first region (\(k_1\)).

The simulation indicates that the ML-estimator of the residual
variance and of the variance of the latent variable were downward
biased (table \ref{SM-tab:biasFactor} in supplementary material
\ref{SM-SM:tableBiasML}), i.e. the estimated variance are too
small. For instance, for \(n\)=20, the average bias was -0.125 for the
residual variance and -0.150 for the variance of the latent variable
(1 is the true value). The proposed correction was partially able to
correct the bias, e.g. for \(n\)=20, the average bias became -0.029 for
the residual variance and to -0.018 for the variance of the latent
variable. Without correction, the inflation of the type 1 error rate
in the robust Wald test was dependent on the coefficient (second row
of \autoref{fig:type1error-robustWald}): \(+0.038\) for \(\nu_2\),
\(+0.063\) for \(\lambda_4\), \(+0.059\) for \(\gamma_1\), and
\(+0.067\) for \(k_1\) when \(n=20\). The complete correction provided
a satisfactory control for \(\nu_2\) and \(\gamma_1\) (type 1 error of
\(0.041\) and \(0.056\), respectively for \(n=20\)) but was slightly
too liberal for \(k_1\) (type 1 error of \(0.074\)) and slightly too
conservative for \(\lambda_4\) (type 1 error of \(0.036\)).

\bigskip

\textbf{Wald test in a LVM with two latent variables (Study C)}: we now add a
new latent variable (\(\eta_{memory}\)) and an additional measurement
per latent variable. The resulting LVM had 36 parameters. The new null
hypotheses tests whether, conditional on the latent variable, there is
a residual covariance between the first two regions (\(\sigma_{12}\))
and whether the first latent variable influences the second (\(b_1\)).

The results were similar to study B: the covariance parameter
\(\sigma_{12}\) did not show any bias, whereas the correction reduced
the bias for the other parameters (table \ref{SM-tab:biasLVM} in
supplementary material \ref{SM-SM:tableBiasML}). The inflation of the
type 1 error was especially noticeable because the model involved more
parameters (third row of \autoref{fig:type1error-Wald}). The bias
correction alone reduced the inflation of the type 1 error rate by
\(0.01\) to \(0.03\). When combined with the Satterthwaite correction,
the resulting procedure was satisfying for \(\gamma_1\) (type 1 error
of \(0.061\) when \(n=20\)), conservative for \(\nu_2\),
\(\lambda_4\), \(b_1\), and \(\sigma_{12}\) (type 1 error of
respectively \(0.032\), \(0.023\), \(0.013\), \(0.024\) when
\(n=20\)), but still liberal for \(k_1\) (type 1 error of \(0.078\)
when \(n=20\)).

\bigskip

\textbf{Distribution of the Wald statistic in small samples}: the simulation
results show that the proposed correction does not always control the
type 1 error rate exactly at the nominal level. This may be due to the
fact that the estimated Wald statistics are not \(t\)-distributed or
that our estimators of the variance and degrees of freedom behave
poorly in small samples. In supplementary material
\ref{SM-SM:validityStudent}, we compare the distribution of the Wald
statistic obtained after correction to the empirical one (obtained by
simulation). We found that our estimators performed well in scenario
A. However, in scenario B and C, the corrected variance was slightly
biased downward and the Satterthwaite estimator of the degrees of
freedom performed poorly for some parameters. The Student's
\(t\)-distribution appeared to be a good approximation of the
empirical distribution except for two types of parameters
(\(b_1\), \(\sigma_{12}\)).

\section{Application of the small sample correction to real data}
\label{applicationSSC}
We now re-run the Wald tests presented at the end of section
\ref{applicationLVM} using the small sample correction developped in
sections \ref{biasCorrection}, \ref{dfEstimation}, and
\ref{Extensions}. \autoref{tab:factor} gives an overview of the statistical
tests performed with or without correction and the associated type 1
error (obtained by simulation).

\begin{table}[!h]
\centering
\begin{tabular}{cccccccccc}
  \toprule
 && \multicolumn{4}{c}{ML} & \multicolumn{4}{c}{ML with correction} \\ 
\cmidrule(lr){3-6} \cmidrule(lr){7-10}
Application & parameter & \(\sigma\) & \(df\) & p-value & type 1 error & \(\sigma\) & \(df\) & p-value & type 1 error \\ \hline
A & $k_1,k_2,k_3$ &  & $\infty$ & \num{1.76e-03} & 10.22\% &  & 49.1 & \num{1.23e-02} & 4.49\% \\ [0.5cm]
B & $\gamma_2$ & 0.026 & $\infty$ & \num{5.11e-03} & 7.41\% & 0.027 & 65.4 & \num{9.03e-03} & 5.59\% \\ 
  & $k_1$ & 0.016 & $\infty$ & \num{7.23e-06} & 6.14\% & 0.017 & 67.5 & \num{4.80e-05} & 4.78\% \\ [0.5cm]
C & $b_1$ & 2.106 & $\infty$ & \num{5.45e-04} & 6.32\% & 2.143 & 13.5 & \num{4.53e-03} & 2.02\% \\ 
  & $b_2$ & 2.355 & $\infty$ & \num{4.21e-03} & 8.50\% & 2.429 & 10.0 & \num{1.96e-02} & 3.71\% \\ 
  & $b_3$ & 2.041 & $\infty$ & \num{7.20e-02} & 8.41\% & 2.130 &  7.7 & \num{1.24e-01} & 3.42\% \\ 
   \bottomrule
\end{tabular}
\caption{Statistical inference without (column ML) and with (column ML with correction) the proposed correction in the three applications (A, B, and C) considered. 
The parameter \(\sigma\) refers to the standard error associated with the estimated parameter and
\(df\) refers to the degrees of freedom of the Wald statistic.
The type 1 error is computed by simulation under the null hypothesis.}
\label{tab:factor}
\end{table}

\bigskip

\textbf{Growth of guinea pigs (Application A)}: the correction we proposed in
 this article shares similarities with the KR correction: we attempt
 to correct the bias of the ML estimator when estimating the variance
 parameters and to use a Student's \(t\)-distribution to model the
 distribution of the Wald statistic. The techniques used however
 differ, e.g., our bias-corrected ML estimator is not identical to the
 REML estimator. Nevertheless, the estimates of the residual variance
 and the variance of the random intercept obtained with our correction
 were similar to those obtained with REML (relative difference <1\%,
 \autoref{tab:mixed}). They were 19\% and 11\% larger than their
 corresponding ML estimates. The corrected test statistic was 4.02
 with a corresponding p-value of 0.0123. Although our method does not
 replicate the results obtained with the KR correction exactly, it
 gives estimates of the same order of magnitude. To further validate
 our correction, we performed the same simulation study as in the
 uncorrected case and found a type 1 error of 0.045 with the proposed
 correction.

\begin{table}[!h]
\centering
\begin{tabular}{rccc}
  \toprule
 & ML & ML with correction & REML \\ 
  \midrule
residual variance & 0.148 & 0.177 & 0.176 \\ 
  variance random intercept & 0.349 & 0.389 & 0.389 \\ 
  statistic & 5.024 & 4.019 & 4.247 \\ 
  degrees of freedom & $\infty$ & 49.11 & 43.59 \\ 
  p-value & 0.0018 & 0.0123 & 0.0102 \\ 
   \bottomrule
\end{tabular}
\caption{Comparison of the ML estimates, 
the proposed correction, 
and the REML estimates (using KR correction, \citet{kenward1997small}) in application A.}
\label{tab:mixed}
\end{table}

\bigskip

\textbf{Serotonin 4 receptor binding and genetic polymorphisms (Application
B)}: compared to the ML-estimates of the variances parameters, the
corrected (non-robust) variance estimates were larger by 23.9\%
(neocortex), 4.5\% (caudate), 4.0\% (putamen), 3.8\% (hippocampus), 3.8\%
(amygdala), and 10.9\% (latent variable). When using cluster robust
standard errors, the corrected p-values were 0.009 for the effect of
the BDNF val66met (+76\% compared to the original p-value) and \(4.8 \;
10^{-5}\) for the effect of 5-HTTLPR (+564\%). The same simulation
study as in the uncorrected case gives, after correction, a type 1
error of 0.056 for the BDNF val66met effect and 0.048 for the 5-HTTLPR
effect. Since the type 1 error estimated by simulations is close to
the nominal level after correction and the corrected p-values are
still below the critical level, this new analysis supports the
conclusions of the original article.

\bigskip

\textbf{Serotonin 4 receptor binding and verbal memory recall (Application
C)}: the bias correction increased the estimates of the variance
parameters by a factor ranging between 4.8\% to 21.2\% (endogenous
variables) and 9.1\% to 10.2\% (latent variables). The corrected
p-values were 0.0045 for \(b_1\) (+731\% compared to the original
p-value), 0.02 for \(b_2\) (+366\%), and 0.12 for \(b_3\) (+72\%). The
same simulation study as in the uncorrected case gives, after
correction, a type 1 of 0.02 for \(b_1\), 0.037 for \(b_2\), and 0.034
for \(b_3\) after correction. This new analysis supports the existence
of an association between serotonin 4 receptor binding and recall of
positive and neutral words (\(b_1\) and \(b_2\)). The parameter
\(b_3\) did not reach significance before correction, where the
testing procedure is liberal, so we should retain the null hypothesis
for \(b_3\). 

\bigskip

In these applications, although the correction did not affect the
conclusion of the statistical tests (when using a significance
threshold of 0.05), the corrected p-values were better calibrated and
therefore better reflected the strength of evidence against the null
hypothesis.

\section{Discussion}
\label{discussion}
Concerns have been raised in the applied scientific literature about
the lack of statistical power and the lack of reliability of studies
involving small samples, e.g., see \citet{button2013power} and
\citet{bakker2012rules}. Compared to \(t\)-tests or linear regressions,
multivariate approaches such as LVMs can be used to increase the power
of testing procedures. They also provide a common framework to test
the hypotheses of the investigator and to assess modeling
assumptions. Although exact tests can be performed on univariate
models, only approximate tests are tractable with LVMs. Using
simulation studies, we performed a detailed investigation of control
of the type 1 error rate when using Wald tests in LVMs with small
samples. The overall conclusion from these simulations was that the type 1
error rate is inflated in small samples. For a sample size of 20, the
type 1 error rate varied between 0.06 to 0.12 for Wald test and
between 0.07 and 0.14 for robust Wald test. The nominal level of 0.05
was reach when the sample size reached 100 to 200, depending on the
type of model parameter. 

\bigskip

We proposed two corrections to obtain a better control of the type 1
error rate: a correction for the bias of the ML-estimator of the
variance parameters and the use of a Student's \(t\)-distribution
instead of a normal distribution to account for the uncertainty in the
estimate of the variance of the model parameters. The proposed
corrections have some desirable features: (1) when combined, they
match the traditional corrections performed in univariate linear
models, (2) they match the uncorrected ML inference in large samples,
(3) they are fast to compute, require no user input, and can be
applied to a large variety of models, and (4) the first correction
reduced the bias of the variance estimates and improved the control of
the type 1 error in all studies. Regarding (3), our implementation had
a very reasonable run time: 75 ms to 475 ms in Study A, 200 ms to 900
ms in Study B, and 1.5s to 6s in Study C. It converged in very few
iterations, except with very small samples and complex LVMs. One
drawback of the proposed testing procedure is that it is not
parametrisation-invariant, meaning that different identifiability
constrains (e.g., setting to 1 a loading in the measurement model or
the variance of the latent variable) may lead to different
p-values. This is a well known issue when using Wald tests, and not a
specificity of our corrections. Alternative test statistics
(e.g. likelihood ratio test) should be considered if this property is
required \citep{larsen2003parametrization}.

\bigskip

A careful inspection of the simulation results showed that using a
\(t\)-distribution to model the distribution of the Wald statistics
was a good approximation for most parameters. We think that the
inexact control of the type 1 error in small samples is mainly due to
the poor performance of our estimator of the degrees of freedom. The
estimation of the standard error could also be improved; indeed our
bias-correction does not completely remove the bias from the estimator
of the variance parameter. A better bias correction may be achieved
using the formula of \citet{cox1968general} for the small sample bias of
the ML estimator. It gives an estimate of the small sample bias up to
\(o_p(n^{-1})\) but involves complex calculations (third order
derivative of the likelihood). The extension of the correction to
robust standard errors could also be improved. Indeed the small sample
correction has been derived assuming that the model was correctly
specified (more precisely that
\(\Esp\left[\trans{\Vxi_i(\Vparam)}\Vxi_i(\Vparam)\right]=\Omega(\Vparam)\)
and \(\Information(\Vparam)^{-1} = \Sigma_{\VparamHat}\)) and the
current approximation for the degrees of freedom does not depend on
the choice of the clusters \(\mathcal{G}_1,\ldots,\mathcal{G}_G\). The
estimation of the degree of freedom will perform poorly when the
clusters contain many observations. We investigated other
approximations (e.g., \cite{pan2002small}) but did not obtain satisfying
results. Finally, we note that the Satterthwaite correction was
derived for the expected information matrix. In theory, a similar
correction could be derived for the observed information matrix, but
it would require more tedious derivations and complexify the software
implementation. Nevertheless, this may be necessary in specific
contexts, e.g., see \citet{savalei2010expected} for a case where the
expected information does not give consistent standard errors (missing
data problems). In our software package implementing the proposed
corrections, we provide a function called \texttt{calibrateType1} that can be
used to assess the type 1 error of the corrected and uncorrected Wald
tests via simulations - under the assumption that the investigator
model is correctly specified. We hope that this will help to detect
inflations in the type 1 error rate and improve the reliability of
studies involving small samples.

\bigskip

As pointed out by one reviewer, alternative estimation techniques such
as instrumental variables (IV, \cite{bollen1996alternative}),
generalized least squares (GLS), and weighted least squares (WLS,
\cite{yuan1997mean}) could compare favorably to ML in small
samples. Although a comprehensive comparison between these estimation
techniques is beyond the scope of the present article, we performed an
additional simulation to compare ML, IV, GLS, and WLS on Study B
under a correctly specified model (see supplementary material
\ref{SM-SM:comparison}). We found that GLS and IV showed an inflation
of the type 1 error in small samples that is similar in magnitude to
ML (uncorrected). WLS failed to estimate the model parameter for
\(n=20\) and \(n=30\); it also had the worst control of the type 1
error. This poor behavior of WLS in small samples is consistent with
the existing litterature \citep{olsson2000performance}. Given the
appealing properties of IV \citep{bollen2007latent}, it would be of
interest to propose a small sample correction for IV estimation.

\section{Acknowledgement}
\label{sec:org687ddfe}
B.O. was supported by the Lundbeck foundation (R231-2016-3236).
This project has received funding from the European Union’s Horizon 2020 research and innovation
programme under the Marie Sklodowska-Curie grant agreement No 746850.

\section*{References}
\label{sec:orgb1010c6}
\begingroup
\renewcommand{\section}[2]{}
\bibliographystyle{apalike}
\bibliography{LVM-smallSampleInference}

\begin{thebibliography}{}

\bibitem[Bakker et~al., 2012]{bakker2012rules}
Bakker, M., van Dijk, A., and Wicherts, J.~M. (2012).
\newblock The rules of the game called psychological science.
\newblock {\em Perspectives on Psychological Science}, 7(6):543--554.

\bibitem[Bentler and Yuan, 1999]{bentler1999structural}
Bentler, P.~M. and Yuan, K.-H. (1999).
\newblock Structural equation modeling with small samples: Test statistics.
\newblock {\em Multivariate behavioral research}, 34(2):181--197.

\bibitem[Bollen, 1996]{bollen1996alternative}
Bollen, K.~A. (1996).
\newblock An alternative two stage least squares (2{S}{L}{S}) estimator for
  latent variable equations.
\newblock {\em Psychometrika}, 61(1):109--121.

\bibitem[Bollen et~al., 2007]{bollen2007latent}
Bollen, K.~A., Kirby, J.~B., Curran, P.~J., Paxton, P.~M., and Chen, F. (2007).
\newblock Latent variable models under misspecification: two-stage least
  squares (2{S}{L}{S}) and maximum likelihood ({M}{L}) estimators.
\newblock {\em Sociological Methods \& Research}, 36(1):48--86.

\bibitem[Button et~al., 2013]{button2013power}
Button, K.~S., Ioannidis, J.~P., Mokrysz, C., Nosek, B.~A., Flint, J.,
  Robinson, E.~S., and Munaf{\`o}, M.~R. (2013).
\newblock Power failure: why small sample size undermines the reliability of
  neuroscience.
\newblock {\em Nature Reviews Neuroscience}, 14(5):365.

\bibitem[Carpenter and Bithell, 2000]{carpenter2000bootstrap}
Carpenter, J. and Bithell, J. (2000).
\newblock Bootstrap confidence intervals: when, which, what? a practical guide
  for medical statisticians.
\newblock {\em Statistics in medicine}, 19(9):1141--1164.

\bibitem[Cox and Snell, 1968]{cox1968general}
Cox, D.~R. and Snell, E.~J. (1968).
\newblock A general definition of residuals.
\newblock {\em Journal of the Royal Statistical Society. Series B
  (Methodological)}, pages 248--275.

\bibitem[da~Cunha-Bang et~al., 2018]{da2018men}
da~Cunha-Bang, S., Fisher, P.~M., Hjordt, L.~V., Perfalk, E., Beliveau, V.,
  Holst, K., and Knudsen, G.~M. (2018).
\newblock Men with high serotonin 1b receptor binding respond to provocations
  with heightened amygdala reactivity.
\newblock {\em NeuroImage}, 166:79--85.

\bibitem[Deen et~al., 2017]{deen2017low}
Deen, M., Hansen, H.~D., Hougaard, A., da~Cunha-Bang, S., N{\o}rgaard, M.,
  Svarer, C., Keller, S.~H., Thomsen, C., Ashina, M., and Knudsen, G.~M.
  (2017).
\newblock Low 5-ht1b receptor binding in the migraine brain: A pet study.
\newblock {\em Cephalalgia}.

\bibitem[Fisher et~al., 2017]{fisher2017bdnf}
Fisher, P., Ozenne, B., Svarer, C., Adamsen, D., Lehel, S., Baar{\'e}, W.,
  Jensen, P., and Knudsen, G. (2017).
\newblock Bdnf val66met association with serotonin transporter binding in
  healthy humans.
\newblock {\em Translational psychiatry}, 7(2):e1029.

\bibitem[Fisher et~al., 2015]{fisher2015bdnf}
Fisher, P.~M., Holst, K.~K., Adamsen, D., Klein, A.~B., Frokjaer, V.~G.,
  Jensen, P.~S., Svarer, C., Gillings, N., Baare, W.~F., Mikkelsen, J.~D.,
  et~al. (2015).
\newblock Bdnf val66met and 5-httlpr polymorphisms predict a human in vivo
  marker for brain serotonin levels.
\newblock {\em Human brain mapping}, 36(1):313--323.

\bibitem[Harville, 1977]{harville1977maximum}
Harville, D.~A. (1977).
\newblock Maximum likelihood approaches to variance component estimation and to
  related problems.
\newblock {\em Journal of the American Statistical Association},
  72(358):320--338.

\bibitem[Herzog et~al., 2007]{herzog2007model}
Herzog, W., Boomsma, A., and Reinecke, S. (2007).
\newblock The model-size effect on traditional and modified tests of covariance
  structures.
\newblock {\em Structural Equation Modeling}, 14(3):361--390.

\bibitem[Holst and Budtz-J{\o}rgensen, 2013]{holst2013linear}
Holst, K.~K. and Budtz-J{\o}rgensen, E. (2013).
\newblock Linear latent variable models: the lava-package.
\newblock {\em Computational Statistics}, 28(4):1385--1452.

\bibitem[Jiang and Yuan, 2017]{jiang2017four}
Jiang, G. and Yuan, K.-H. (2017).
\newblock Four new corrected statistics for {S}{E}{M} with small samples and
  nonnormally distributed data.
\newblock {\em Structural Equation Modeling: A Multidisciplinary Journal},
  24(4):479--494.

\bibitem[Kauermann and Carroll, 2001]{goran2001note}
Kauermann, G. and Carroll, R.~J. (2001).
\newblock A note on the efficiency of sandwich covariance matrix estimation.
\newblock {\em Journal of the American Statistical Association},
  96(456):1387--1396.

\bibitem[Kenward and Roger, 1997]{kenward1997small}
Kenward, M.~G. and Roger, J.~H. (1997).
\newblock Small sample inference for fixed effects from restricted maximum
  likelihood.
\newblock {\em Biometrics}, pages 983--997.

\bibitem[Kuznetsova et~al., 2017]{kuznetsova2017lmertest}
Kuznetsova, A., Brockhoff, P.~B., and Christensen, R. H.~B. (2017).
\newblock lmertest package: tests in linear mixed effects models.
\newblock {\em Journal of Statistical Software}, 82(13).

\bibitem[Larsen et~al., 2003]{larsen2003parametrization}
Larsen, P.~V., Jupp, P., et~al. (2003).
\newblock Parametrization-invariant wald tests.
\newblock {\em Bernoulli}, 9(1):167--182.

\bibitem[Maydeu-Olivares, 2017]{maydeu2017maximum}
Maydeu-Olivares, A. (2017).
\newblock Maximum likelihood estimation of structural equation models for
  continuous data: Standard errors and goodness of fit.
\newblock {\em Structural Equation Modeling: A Multidisciplinary Journal},
  24(3):383--394.

\bibitem[McNeish, 2016]{mcneish2016using}
McNeish, D. (2016).
\newblock On using bayesian methods to address small sample problems.
\newblock {\em Structural Equation Modeling: A Multidisciplinary Journal},
  23(5):750--773.

\bibitem[Muth\'{e}n and Muth\'{e}n, 2017]{Mplus}
Muth\'{e}n, L.~K. and Muth\'{e}n, B.~O. (2017).
\newblock {Mplus} user's guide. eighth edition.

\bibitem[Olsson et~al., 2000]{olsson2000performance}
Olsson, U.~H., Foss, T., Troye, S.~V., and Howell, R.~D. (2000).
\newblock The performance of {M}{L}, {G}{L}{S}, and {W}{L}{S} estimation in
  structural equation modeling under conditions of misspecification and
  nonnormality.
\newblock {\em Structural equation modeling}, 7(4):557--595.

\bibitem[Pan and Wall, 2002]{pan2002small}
Pan, W. and Wall, M.~M. (2002).
\newblock Small-sample adjustments in using the sandwich variance estimator in
  generalized estimating equations.
\newblock {\em Statistics in medicine}, 21(10):1429--1441.

\bibitem[Parr, 1983]{parr1983note}
Parr, W.~C. (1983).
\newblock A note on the jackknife, the bootstrap and the delta method
  estimators of bias and variance.
\newblock {\em Biometrika}, 70(3):719--722.

\bibitem[Pek and Wu, 2015]{pek2015profile}
Pek, J. and Wu, H. (2015).
\newblock Profile likelihood-based confidence intervals and regions for
  structural equation models.
\newblock {\em Psychometrika}, 80(4):1123--1145.

\bibitem[Perfalk et~al., 2017]{perfalk2017testosterone}
Perfalk, E., da~Cunha-Bang, S., Holst, K.~K., Keller, S., Svarer, C., Knudsen,
  G.~M., and Frokjaer, V.~G. (2017).
\newblock Testosterone levels in healthy men correlate negatively with
  serotonin 4 receptor binding.
\newblock {\em Psychoneuroendocrinology}, 81:22--28.

\bibitem[Rosseel, 2012]{lavaan}
Rosseel, Y. (2012).
\newblock {lavaan}: An {R} package for structural equation modeling.
\newblock {\em Journal of Statistical Software}, 48(2):1--36.

\bibitem[Satorra and Bentler, 1994]{satorra1994corrections}
Satorra, A. and Bentler, P.~M. (1994).
\newblock Corrections to test statistics and standard errors in covariance
  structure analysis.
\newblock {\em Latent Varaibles Analysis: Applications to Developmental
  Research}.

\bibitem[Savalei, 2010]{savalei2010expected}
Savalei, V. (2010).
\newblock Expected versus observed information in {S}{E}{M} with incomplete
  normal and nonnormal data.
\newblock {\em Psychological methods}, 15(4):352.

\bibitem[Stenb{\ae}k et~al., 2017]{stenbaek2017brain}
Stenb{\ae}k, D.~S., Fisher, P.~M., Ozenne, B., Andersen, E., Hjordt, L.~V.,
  McMahon, B., Hasselbalch, S.~G., Frokjaer, V.~G., and Knudsen, G.~M. (2017).
\newblock Brain serotonin 4 receptor binding is inversely associated with
  verbal memory recall.
\newblock {\em Brain and behavior}, 7(4).

\bibitem[Wei et~al., 1998]{wei1998generalized}
Wei, B.-C., Hu, Y.-Q., and Fung, W.-K. (1998).
\newblock Generalized leverage and its applications.
\newblock {\em Scandinavian Journal of statistics}, 25(1):25--37.

\bibitem[White, 1982]{white1982maximum}
White, H. (1982).
\newblock Maximum likelihood estimation of misspecified models.
\newblock {\em Econometrica: Journal of the Econometric Society}, pages 1--25.

\bibitem[Wu and Lin, 2016]{wu2016scaled}
Wu, H. and Lin, J. (2016).
\newblock A scaled f distribution as an approximation to the distribution of
  test statistics in covariance structure analysis.
\newblock {\em Structural Equation Modeling: A Multidisciplinary Journal},
  23(3):409--421.

\bibitem[Yuan and Bentler, 1997]{yuan1997mean}
Yuan, K.-H. and Bentler, P.~M. (1997).
\newblock Mean and covariance structure analysis: Theoretical and practical
  improvements.
\newblock {\em Journal of the American statistical association},
  92(438):767--774.

\end{thebibliography}


\begin{thebibliography}{}

\end{thebibliography}
\endgroup
\end{document}


\maketitle
\appendix
\renewcommand{\thesection}{\Alph{section}}
\renewcommand{\thesubsection}{\Alph{section}.\arabic{subsection}}

\section{Likelihood and derivatives in LVMs}
\label{sec:org8e4f6a7}

In this section, we show the expression of the likelihood, its first
two derivatives, the information matrix, and the first derivative of the
information matrix.

\subsection{Likelihood}
\label{SM:likelihood}
At the individual level, the measurement and structural models can be written:
\begin{align*}
\VY_i &= \nu + \Veta_i \Lambda + \VX_i K + \Vvarepsilon_i \\
\Veta_i &= \alpha + \Veta_i B + \VX_i \Gamma + \boldsymbol{\zeta}_i 
\end{align*}
\begin{tabular}{lll}
with & \(\Sigma_{\epsilon}\)   &the variance-covariance matrix of the residuals \(\Vvarepsilon_i\)\\
     & \(\Sigma_{\zeta}\) & the variance-covariance matrix of the residuals \(\boldsymbol{\zeta}_i\). \\
\end{tabular}

\bigskip

By combining the previous equations, we can get an expression for
\(\VY_i\) that does not depend on \(\Veta_i\):
\begin{align*}
\VY_i &= \nu + \left(\boldsymbol{\zeta}_i + \alpha + \VX_i \Gamma \right) (I-B)^{-1} \Lambda + \VX_i K + \Vvarepsilon_i 
\end{align*}
Since \(\Var[Ax] = A \Var[x] \trans{A}\) we have \(\Var[xA] =
\trans{A} \Var[x] A\), we have the following expressions for the
conditional mean and variance of \(\VY_i\):
\begin{align*}
 \Vmu(\Vparam,\VX_i) &= E[\VY_i|\VX_i] = \nu + (\alpha + \VX_i \Gamma) (1-B)^{-1} \Lambda + \VX_i K \\
\Omega(\Vparam) &= Var[\VY_i|\VX_i] = \Lambda^t (1-B)^{-t}  \Sigma_{\zeta} (1-B)^{-1} \Lambda + \Sigma_{\varepsilon} 
\end{align*}

\bigskip

where \(\Vparam\) is the collection of all parameters. The
log-likelihood can be written:
\begin{align*}
l(\Vparam|\VY,\VX) &= \sum_{i=1}^n l(\Vparam|\VY_i,\VX_i) \\
&= \sum_{i=1}^{n} - \frac{p}{2} log(2\pi) - \frac{1}{2} log|\Omega(\Vparam)| - \frac{1}{2} (\VY_i-\Vmu(\Vparam,\VX_i)) \Omega(\Vparam)^{-1} \trans{(\VY_i-\Vmu(\Vparam,\VX_i))}
\end{align*}

\subsection{First derivative: score}
\label{SM:score}
The individual score is obtained by derivating the log-likelihood:
\begin{align*}
   \Score(\param|\VY_i,\VX_i) =& \dpartial{l_i(\Vparam|\VY_i,\VX_i)}{\param}\\
 =& - \frac{1}{2} tr\left(\Omega(\Vparam)^{-1} \dpartial{\Omega(\Vparam)}{\param}\right) \\
 &+  \dpartial{\Vmu(\Vparam,\VX_i)}{\param} \Omega(\Vparam)^{-1} \trans{(\VY_i-\Vmu(\Vparam,\VX_i))} \\
 &+ \frac{1}{2} (\VY_i-\Vmu(\Vparam,\VX_i)) \Omega(\Vparam)^{-1} \dpartial{\Omega(\Vparam)}{\param} \Omega(\Vparam)^{-1} \trans{(\VY_i-\Vmu(\Vparam,\VX_i))}
\end{align*}

\subsection{Second derivative: Hessian and expected information}
\label{SM:Information}
The individual Hessian is obtained by derivating twice the
log-likelihood:
\begin{align*}
   \Hessian_i(\param,\param') =& -\frac{1}{2} tr\left(-\Omega(\Vparam)^{-1} \dpartial{\Omega(\Vparam)}{\param'} \Omega(\Vparam)^{-1} \frac{\partial \Omega(\Vparam)}{\partial \param} + \Omega(\Vparam)^{-1} \frac{\partial^2 \Omega(\Vparam)}{\partial \param \partial \param'}\right) \\
 &+  \frac{\partial^2 \Vmu(\Vparam,\VX_i)}{\partial \param \partial \param'} \Omega(\Vparam)^{-1} \trans{(\VY_i-\Vmu(\Vparam,\VX_i))} \\
 &-  \dpartial{\Vmu(\Vparam,\VX_i)}{\param} \Omega(\Vparam)^{-1} \dpartial{\Omega(\Vparam)}{\param'} \Omega(\Vparam)^{-1} \trans{(\VY_i-\Vmu(\Vparam,\VX_i))} \\
 &-  \dpartial{\Vmu(\Vparam,\VX_i)}{\param} \Omega(\Vparam)^{-1} \trans{\dpartial{\Vmu(\Vparam,\VX_i)}{\param'}} \\
 &-  \dpartial{\Vmu(\Vparam,\VX_i)}{\param'} \Omega(\Vparam)^{-1} \dpartial{\Omega(\Vparam)}{\param} \Omega(\Vparam)^{-1} \trans{(\VY_i-\Vmu(\Vparam,\VX_i))}  \\
 &-  (\VY_i-\Vmu(\Vparam,\VX_i)) \Omega(\Vparam)^{-1} \dpartial{\Omega(\Vparam)}{\param'} \Omega(\Vparam)^{-1} \frac{\partial \Omega(\Vparam)}{\partial \param} \Omega(\Vparam)^{-1} \trans{(\VY_i-\Vmu(\Vparam,\VX_i))} \\
 &+ \frac{1}{2} (\VY_i-\Vmu(\Vparam,\VX_i)) \Omega(\Vparam)^{-1} \frac{\partial^2 \Omega(\Vparam)}{\partial \param \partial \param'} \Omega(\Vparam)^{-1} \trans{(\VY_i-\Vmu(\Vparam,\VX_i))} \\
\end{align*}

\clearpage

Using that \(\Vmu(\param,\VX_i)\) and \(\Omega(\Vparam)\) are deterministic quantities,
we can then take the expectation to obtain:
\begin{align*}
\Esp\left[\Hessian_i(\param,\param')\right] =& -\frac{1}{2} tr\left(-\Omega(\Vparam)^{-1} \dpartial{\Omega(\Vparam)}{\param'} \Omega(\Vparam)^{-1} \frac{\partial \Omega(\Vparam)}{\partial \Vparam} + \Omega(\param)^{-1} \frac{\partial^2 \Omega(\Vparam)}{\partial \param \partial \param'}\right) \\
 &+  \frac{\partial^2 \Vmu(\Vparam,\VX_i)}{\partial \param \partial \param'} \Omega(\Vparam)^{-1} \Ccancelto[red]{0}{\Esp\left[\trans{(\VY_i-\Vmu(\Vparam,\VX_i))}\right]} \\
 &-  \dpartial{\Vmu(\Vparam,\VX_i)}{\param} \Omega(\Vparam)^{-1} \dpartial{\Omega(\Vparam)}{\param'} \Omega(\Vparam)^{-1} \Ccancelto[red]{0}{\Esp\left[\trans{(\VY_i-\Vmu(\Vparam,\VX_i))}\right]} \\
 &-  \dpartial{\Vmu(\Vparam,\VX_i)}{\param} \Omega(\Vparam)^{-1} \trans{\dpartial{\Vmu(\Vparam)}{\param'}} \\
 &-  \dpartial{\Vmu(\Vparam,\VX_i)}{\param'} \Omega(\Vparam)^{-1} \dpartial{\Omega(\Vparam)}{\param} \Omega(\Vparam)^{-1} \Ccancelto[red]{0}{\Esp\left[\trans{(\VY_i-\Vmu(\Vparam,\VX_i))}\right]}  \\
 &-  \Esp\left[(\VY_i-\Vmu(\Vparam,\VX_i)) \Omega(\Vparam)^{-1} \dpartial{\Omega(\Vparam)}{\param'} \Omega(\Vparam)^{-1} \frac{\partial \Omega(\Vparam)}{\partial \param} \Omega(\Vparam)^{-1} \trans{(\VY_i-\Vmu(\Vparam,\VX_i))}\right] \\
 &+ \Esp \left[\frac{1}{2} (\VY_i-\Vmu(\Vparam,\VX_i)) \Omega(\Vparam)^{-1} \frac{\partial^2 \Omega(\Vparam)}{\partial \param \partial \param'} \Omega(\Vparam)^{-1} \trans{(\VY_i-\Vmu(\Vparam,\VX_i))}\right] \\
\end{align*}

The last two expectations can be re-written using that \(\Esp[\trans{x}Ax] = tr\left(A\Var[x]\right)+\trans{\Esp[x]}A\Esp[x]\):
\begin{align*}
\Esp\left[\Hessian_i(\param,\param')\right] =& -\frac{1}{2} tr\left(-\Omega(\Vparam)^{-1} \dpartial{\Omega(\Vparam)}{\param'} \Omega(\Vparam)^{-1} \frac{\partial \Omega(\Vparam)}{\partial \param} + \Ccancel[red]{\Omega(\Vparam)^{-1} \frac{\partial^2 \Omega(\Vparam)}{\partial \param \partial \param'}}\right) \\
 &-  \frac{\partial \Vmu(\Vparam,\VX_i)}{\partial \param} \Omega(\Vparam)^{-1} \trans{\frac{\partial \Vmu(\Vparam,\VX_i)}{\partial \param'}} \\
 &- tr\left(\Omega(\Vparam)^{-1} \frac{\partial \Omega(\Vparam)}{\partial \param'} \Omega(\Vparam)^{-1} \frac{\partial \Omega(\Vparam)}{\partial \param} \Omega(\Vparam)^{-1} \trans{\left(\Var\left[\VY_i-\Vmu(\Vparam,\VX_i)\right]\right)} \right) \\
 &+ \Ccancel[red]{\frac{1}{2} tr\left( \Omega(\Vparam)^{-1} \frac{\partial^2 \Omega(\Vparam)}{\partial \param \partial \param'} \Ccancel[blue]{\Omega(\Vparam)^{-1}} \Ccancel[blue]{\trans{\left(\Var\left[\VY_i-\Vmu(\Vparam,\VX_i)\right]\right)}} \right)} \\
\end{align*}
where we have used that \(\Var\left[\VY_i-\Vmu(\Vparam,\VX_i)\right] =
\Var\left[\VY_i|\VX_i\right] = \Omega(\Vparam)\). Finally we get:
\begin{align*}
\Esp\left[\Hessian_i(\param,\param')\right] =& -\frac{1}{2} tr\left(\Omega(\Vparam)^{-1} \dpartial{\Omega(\Vparam)}{\param'} \Omega(\Vparam)^{-1} \dpartial{\Omega(\Vparam)}{\param}\right) \\
 &-  \dpartial{\Vmu(\Vparam,\VX_i)}{\param} \Omega(\Vparam)^{-1} \trans{\dpartial{\Vmu(\Vparam,\VX_i)}{\param'}} \\
\end{align*}
So we can deduce from the previous equation the expected information matrix:
\begin{align*}
\Information(\param,\param') =& \frac{n}{2} tr\left(\Omega(\Vparam)^{-1} \dpartial{\Omega(\Vparam)}{\param'} \Omega(\Vparam)^{-1} \frac{\partial \Omega(\Vparam)}{\partial \param}\right) 
 + \sum_{i=1}^n \dpartial{\Vmu(\Vparam,\VX_i)}{\param} \Omega(\Vparam)^{-1} \trans{\dpartial{\Vmu(\Vparam,\VX_i)}{\param'}}
\end{align*}

\subsection{First derivatives of the information matrix}
\label{SM:dInformation}
\begin{align*}
\frac{\partial \Information(\param,\param')}{\partial \param''} 
=& - \frac{n}{2} tr\left(\Omega(\Vparam)^{-1} \frac{\partial \Omega(\Vparam)}{\partial \param''} \Omega(\Vparam)^{-1} \frac{\partial \Omega(\Vparam)}{\partial \param'} \Omega(\Vparam)^{-1} \frac{\partial \Omega(\Vparam)}{\partial \param}\right) \\
& + \frac{n}{2} tr\left( \Omega(\Vparam)^{-1} \frac{\partial^2 \Omega(\Vparam)}{\partial\param''\partial\param'} \Omega(\Vparam)^{-1} \frac{\partial \Omega(\Vparam)}{\partial \param}\right) \\
& - \frac{n}{2} tr\left(\Omega(\Vparam)^{-1} \frac{\partial \Omega(\Vparam)}{\partial \param'} \Omega(\Vparam)^{-1} \frac{\partial \Omega(\Vparam)}{\partial \param''} \Omega(\Vparam)^{-1} \frac{\partial \Omega(\Vparam)}{\partial \param}\right) \\
& + \frac{n}{2} tr\left( \Omega(\Vparam)^{-1} \frac{\partial \Omega(\Vparam)}{\partial\param'} \Omega(\Vparam)^{-1} \frac{\partial^2 \Omega(\Vparam)}{\partial \param'' \partial \param}\right) \\
& + \sum_{i=1}^n \frac{\partial^2 \Vmu(\Vparam,\VX_i)}{\partial\param\partial\param''} \Omega(\Vparam)^{-1} \trans{\dpartial{\Vmu(\Vparam,\VX_i)}{\param'}} \\
& + \sum_{i=1}^n \frac{\partial \Vmu(\Vparam,\VX_i)}{\partial \param} \Omega(\Vparam)^{-1} \trans{\ddpartial{\Vmu(\Vparam,\VX_i)}{\param'}{\param''}} \\
& - \sum_{i=1}^n \frac{\partial \Vmu(\Vparam,\VX_i)}{\partial \param} \Omega(\Vparam)^{-1} \frac{\partial \Omega(\Vparam)}{\partial \param''} \Omega(\Vparam)^{-1} \trans{\dpartial{\Vmu(\Vparam,\VX_i)}{\param'}} \\
\end{align*}

\clearpage

\section{Residuals in LVMs}
\label{SM:varResiduals}
In this section we will derive the first order bias of the variance of
the observed residuals:
\begin{align*}
\Esp \left[\trans{\Vxi_i(\VparamHat)} \Vxi_i(\VparamHat) \right]
&= \Omega(\Vparam) - \trans{\dpartial{\Vmu(\Vparam,\VX_i)}{\Vparam}}  \Sigma_{\VparamHat} \dpartial{\Vmu(\Vparam,\VX_i)}{\Vparam} + o_p(n^{-1}) 
\end{align*}
and then introduce corrected residuals that have appropriate variance
(up to the first order). Finally we provide a more explicit formula
for the effective sample size \(\Vn^{c} = \sum_{i=1}^n
\dpartial{\Vxi_i(\VparamHat)}{\VY_i}\).

\bigskip

As in the main manuscript, we will denote by \(\Vxi_i(\Vparam) = \VY_i -
\Vmu(\Vparam,\VX_i)\) the (theoretical) residuals for subject \(i\), and
by \(\Vxi_i(\VparamHat) = \VY_i - \Vmu(\VparamHat,\VX_i)\) the
observed residuals for subject \(i\). We will assume that the
variance-covariance matrix of the (theoretical) residuals equals
\(\Omega(\Vparam)\) and
\(\Sigma_{\VparamHat}=\Information(\Vparam)^{-1}\). We use the
convention that \(\Information(\Vparam)\) is the information matrix
relative to the whole sample and we will denote by
\(\Information_1(\Vparam)=\Esp[-\Hessian_i(\Vparam)]\) the information matrix for a single
observation (for iid observations \(\Information(\Vparam) =
n\Information_1(\Vparam)\)).

\subsection{Variance of the observed residuals}
\label{sec:orgadf0a38}
\subsubsection{Lemma}
\label{sec:org3b847a6}
We will use the following lemmas:

\begin{lemma} \label{lemma:iid2} (adapted from formula 2.6 in \cite{vecchia2012higher}) \\
Denoting \(\VZ_j = (\VY_j,\VX_j)\), \(\Hessian_j(\Vparam)\) the Hessian
relative to observation \(j\), and \(\Score^{(p)}(\VZ)\) the \(p\)-th
element of the score vector, the second order decomposition of the ML
estimator can be written:
\begin{align*}
\VparamHat - \Vparam &= \frac{1}{n} \sum_{j=1}^n \varphi_1(\VZ_j,\Vparam) + \frac{1}{2n^2} \sum_{j=1}^n \sum_{k=1}^n \varphi_2(\VZ_j,\VZ_k,\Vparam)  + o_p(n^{-1})  \\
\text{with } \varphi_1(\VZ_j,\Vparam) &= \Score_j(\Vparam|\VY_j,\VX_j) \Information_1(\Vparam)^{-1} \\
\text{and } \varphi_2(\VZ_j,\VZ_k,\Vparam) &= \varphi_1(\VZ_j,\Vparam) + \varphi_1(\VZ_k,\Vparam) + \Information_1(\Vparam)^{-1} \gamma(\VZ_j,\VZ_k,\Vparam) \\
& \qquad + \Information_1^{-1}(\Vparam) (\Hessian_k(\Vparam)\varphi_1(\VZ_j,\Vparam)+\Hessian_j(\Vparam)\varphi_1(\VZ_k,\Vparam))
\end{align*}
where
\begin{align*}
\trans{\gamma(\VZ_j,\VZ_k,\Vparam)} = \left(\begin{array}{c}
\trans{\varphi_1(\VZ_k,\Vparam)} \Esp\left[\ddpartial{\Score^{(1)}(\VZ)}{\param}{\param'}\right] \varphi_1(\VZ_j,\Vparam) \\
\ldots \\
\trans{\varphi_1(\VZ_k,\Vparam)} \Esp\left[\ddpartial{\Score^{(p)}(\VZ)}{\param}{\param'}\right] \varphi_1(\VZ_j,\Vparam) \\
 \end{array}\right)
\end{align*}
\end{lemma}

\bigskip

\begin{lemma} \label{lemma:termB}
In a LVM, we have for the \(i\)-th observation:
\begin{align*}
\Esp \left[\trans{\Vxi_i(\Vparam)} (\VparamHat - \Vparam)\right] &= \frac{1}{n} \Esp \left[ 
\trans{\Vxi_i(\Vparam)} \Score_i(\Vparam|\VY_i,\VX_i)
\right] \Information(\Vparam)^{-1} + o_p(n^{-1}) 
\end{align*}
\end{lemma}

\begin{proof} From lemma \ref{lemma:iid2} we get that:
\begin{align*}
\Esp \left[\trans{\Vxi(\Vparam)}_i (\VparamHat - \Vparam)\right] 
&= \frac{1}{n} \sum_{j=1}^n \Esp \left[ \trans{\Vxi_i(\Vparam)}\varphi_1(\VZ_j,\Vparam)\right] 
+ \frac{1}{2n^2} \sum_{j=1}^n \sum_{k=1}^n \Esp \left[ \trans{\Vxi_i(\Vparam)}\varphi_2(\VZ_j,\VZ_k,\Vparam)\right] 
+ o_p(n^{-1}) 
\end{align*}
Since
\(\Esp\left[\Vxi_i(\Vparam)\right]=0\)
and we have iid observations, this simplifies into:
\begin{align*}
B =& \frac{1}{n} \Esp \left[\trans{\Vxi_i(\Vparam)} \varphi_1(\VZ_i,\Vparam)\right] + \frac{1}{2n^2} \sum_{j=1}^n \Esp \left[\trans{\Vxi_i(\Vparam)} (\varphi_2(\VZ_i,\VZ_j,\Vparam) + \varphi_2(\VZ_j,\VZ_i,\Vparam))\right]
+ o_p(n^{-1})
\end{align*}
If \(i \neq j\), we get from \(\Esp[\varphi_1(\VZ_j,\Vparam)]=0\) that:
\begin{align*}
\Esp \left[ \trans{\Vxi_i(\Vparam)}\varphi_2(\VZ_i,\VZ_j,\Vparam) \right] 
&= \Esp[\trans{\Vxi_i(\Vparam)} \varphi_1(\VZ_i,\Vparam)] + \Esp[\trans{\Vxi_i(\Vparam)} \Information^{-1}(\Vparam) \Hessian_j(\Vparam) \varphi_1(\VZ_i,\Vparam)] \\
&= \Esp[\trans{\Vxi_i(\Vparam)} \varphi_1(\VZ_i,\Vparam)] + \Esp\left[\trans{\Vxi_i(\Vparam)} \Information^{-1}(\Vparam) \Esp\left[\Hessian_j(\Vparam)\right] \varphi_1(\VZ_i,\Vparam)\right] = 0 
\end{align*}
where the last equality follows from \(\Esp[\Hessian_j(\Vparam)] =
-\Information(\Vparam)\). By symmetry \(\Esp \left[
\trans{\Vxi_i(\Vparam)}\varphi_2(\VZ_j,\VZ_i,\Vparam) \right] = 0\) when
\(i \neq j\). Hence:
\begin{align*}
B =& \frac{1}{n} \Esp \left[\trans{\Vxi_i(\Vparam)} \varphi_1(\VZ_i,\Vparam)\right] + \frac{1}{2n^2} \Esp \left[\trans{\Vxi_i(\Vparam)} (\varphi_2(\VZ_i,\VZ_i,\Vparam) + \varphi_2(\VZ_i,\VZ_i,\Vparam))\right]
+ o_p(n^{-1}) \\
&= \frac{1}{n} \Esp \left[\trans{\Vxi_i(\Vparam)} \varphi_1(\VZ_i,\Vparam)\right] + o_p(n^{-1}) 
\end{align*}
because \(\Vxi_i(\Vparam)\) and \(\varphi_2(\VZ_i,\VZ_i,\Vparam)\) are
\(O_p(1)\).
\end{proof}

\begin{lemma} \label{lemma:termC}
In a LVM, we have for the \(i\)-th observation:
\begin{align*}
C =\Esp \left[\trans{\Vxi_i(\Vparam)} (\VparamHat - \Vparam) \frac{\partial^2 \Vmu(\Vparam,\VX_i)}{(\partial \Vparam)^2} \trans{(\VparamHat - \Vparam)}\right] &=  o_p(n^{-1}) 
\end{align*}
\end{lemma}

\begin{proof} From lemma \ref{lemma:iid2} we get that:
\begin{align*}
C = & \frac{1}{n^2} \sum_{j=1}^n \sum_{j'=1}^n \underbrace{\Esp \left[ \trans{\Vxi_i(\Vparam)}\varphi_1(\VZ_j,\Vparam) \frac{\partial^2 \Vmu(\Vparam,\VX_i)}{(\partial \Vparam)^2} \trans{\varphi_1(\VZ_{j'},\Vparam)} \right]}_{C_1} \\
&+\frac{1}{2n^3} \sum_{j=1}^n \sum_{j'=1}^n \sum_{k'=1}^n \underbrace{\Esp \left[ \trans{\Vxi_i(\Vparam)}\varphi_1(\VZ_j,\Vparam) \frac{\partial^2 \Vmu(\Vparam,\VX_i)}{(\partial \Vparam)^2} \trans{\varphi_2(\VZ_{j'},\VZ_{k'},\Vparam)} \right]}_{C_2}
+ o_p(n^{-1}) 
\end{align*}
because each second order term (i.e. the term involving \(\varphi_2\)
) is \(o_p(n^{-\half})\) so their product is \(o_p(n^{-1})\). Because
\(\Esp[\Vxi_i(\Vparam)]=\Esp[\varphi_1(\VZ_j,\Vparam)]=0\) and using that
the observations are iid, \(C_1\) is only non-0 when
\(i=j=j'\). Similarly, \(C_2\) is only non-0 when \(i=j=j'\) or
\(i=j=k'\). Therefore:
\begin{align*}
C = & \frac{1}{n^2} \Esp \left[ \trans{\Vxi_i(\Vparam)}\varphi_1(\VZ_i,\Vparam) \frac{\partial^2 \Vmu(\Vparam,\VX_i)}{(\partial \Vparam)^2} \trans{\varphi_1(\VZ_{i},\Vparam)} \right] \\
&\frac{1}{2n^3} \sum_{j=1}^n \Esp \left[ \trans{\Vxi_i(\Vparam)}\varphi_1(\VZ_i,\Vparam) \frac{\partial^2 \Vmu(\Vparam,\VX_i)}{(\partial \Vparam)^2} \left( \trans{\varphi_2(\VZ_{i},\VZ_{j},\Vparam)} + \trans{\varphi_2(\VZ_{j},\VZ_{i},\Vparam)}\right) \right]
+ o_p(n^{-1}) \\
&= O_p(n^2) + O_p(n^2) + o_p(n^{-1}) = o_p(n^{-1})
\end{align*}
where we have used that \(\Vxi_i(\Vparam)\),
\(\varphi_1(\VZ_i,\Vparam)\), \(\frac{\partial^2
\Vmu(\Vparam,\VX_i)}{(\partial \Vparam)^2}\), and
\(\varphi_2(\VZ_{i},\VZ_{j},\Vparam)\) are \(O_p(1)\).

\end{proof}
\subsubsection{Proof of the main result}
\label{sec:org9df4fd1}
To compute the variance of the observed residuals, we first express
 the observed residuals as a function of the theoretical residuals
 using a Taylor expansion:
\begin{align*}
\Vmu(\VparamHat,\VX_i) &= \Vmu(\Vparam,\VX_i) + (\VparamHat - \Vparam)\dpartial{\Vmu(\Vparam,\VX_i)}{\Vparam} + (\VparamHat - \Vparam) \frac{\partial^2 \Vmu(\Vparam,\VX_i)}{(\partial \Vparam)^2} \trans{(\VparamHat - \Vparam)}  + O_p(||\VparamHat - \Vparam||^3) \\
\VY_i - \Vmu(\VparamHat,\VX_i) &= \VY_i - \Vmu(\Vparam,\VX_i) - (\VparamHat - \Vparam)\dpartial{\Vmu(\Vparam,\VX_i)}{\Vparam} - (\VparamHat - \Vparam) \frac{\partial^2 \Vmu(\Vparam,\VX_i)}{(\partial \Vparam)^2} \trans{(\VparamHat - \Vparam)} + O_p(||\VparamHat - \Vparam||^3) \\
\Vxi_i(\VparamHat) &= \Vxi_i(\Vparam) - (\VparamHat - \Vparam) \dpartial{\Vmu(\Vparam,\VX_i)}{\Vparam} - (\VparamHat - \Vparam) \frac{\partial^2 \Vmu(\Vparam,\VX_i)}{(\partial \Vparam)^2} \trans{(\VparamHat - \Vparam)} + O_p(||\VparamHat - \Vparam||^3)
\end{align*}
where \(||v||\) denotes the euclidean norm of the vector
\(v\). Using that \(\dpartial{\Vmu(\Vparam,\VX_i)}{\Vparam}\) is non-stochastic, we get:
\begin{align*}
\Esp \left[\trans{\Vxi_i(\VparamHat)} \Vxi_i(\VparamHat) \right]
&= \textcolor{red}{\Esp \left[ \trans{\Vxi_i(\Vparam)}\Vxi_i(\Vparam) \right]}
- \textcolor{blue}{\Esp \left[\trans{\Vxi_i(\Vparam)} (\VparamHat - \Vparam)\right]} \dpartial{\Vmu(\Vparam,\VX_i)}{\Vparam}  
- \textcolor{gray}{ \Esp \left[\trans{\Vxi_i(\Vparam)} (\VparamHat - \Vparam) \frac{\partial^2 \Vmu(\Vparam,\VX_i)}{(\partial \Vparam)^2} \trans{(\VparamHat - \Vparam)}  \right]}  \\
&- \trans{\dpartial{\Vmu(\Vparam,\VX_i)}{\Vparam}}  \textcolor{blue}{ \Esp \left[\trans{(\VparamHat - \Vparam)} \Vxi_i(\Vparam) \right]}
+ \trans{\dpartial{\Vmu(\Vparam,\VX_i)}{\Vparam}} \textcolor{green}{ \Esp \left[  \trans{(\VparamHat - \Vparam)} (\VparamHat - \Vparam) \right]} \dpartial{\Vmu(\Vparam,\VX_i)}{\Vparam}  \\
& - \textcolor{gray}{ \Esp \left[(\VparamHat - \Vparam) \frac{\partial^2 \Vmu(\Vparam,\VX_i)}{(\partial \Vparam)^2} \trans{(\VparamHat - \Vparam)} \Vxi_i(\Vparam)  \right]}  
 + O_p(||\VparamHat - \Vparam||^3) \\
&= \textcolor{red}{A} - \textcolor{blue}{B} \dpartial{\Vmu(\Vparam,\VX_i)}{\Vparam} - \textcolor{gray}{C} + \trans{\dpartial{\Vmu(\Vparam,\VX_i)}{\Vparam}} \textcolor{blue}{\trans{B}} + \trans{\dpartial{\Vmu(\Vparam,\VX_i)}{\Vparam}} \textcolor{green}{D} \dpartial{\Vmu(\Vparam,\VX_i)}{\Vparam} - \textcolor{gray}{\trans{C}}
+ O_p(||\VparamHat - \Vparam||^3)
\end{align*}
We now study each of the term seperately:

\bigskip
\begin{itemize}
\item \textbf{term A}: \(\Esp \left[ \trans{\Vxi_i(\Vparam)}\Vxi_i(\Vparam) \right] = \Omega(\Vparam)\)

\item \textbf{term B}: from lemma \ref{lemma:termB} we get that:
\end{itemize}
\begin{align*}
B &= \Esp \left[ 
\trans{\Vxi_i(\Vparam)} \Score_i(\Vparam|\VY_i,\VX_i) \right] \Information(\Vparam)^{-1} + o_p(n^{-1}) 
\end{align*}
Moreover, from the score equations (appendix \ref{SM:Information}), we
have:
\begin{align*}
&   \Score_i(\param|\VY_i,\VX_i)  \\
=& - \frac{1}{2} tr\left(\Omega(\Vparam)^{-1} \dpartial{\Omega(\Vparam)}{\param}\right) 
 +  \dpartial{\Vmu(\Vparam,\VX_i)}{\param} \Omega(\Vparam)^{-1} \trans{\Vxi_i(\Vparam)} 
 + \frac{1}{2} \Vxi_i(\Vparam) \Omega(\Vparam)^{-1} \dpartial{\Omega(\Vparam)}{\param} \Omega(\Vparam)^{-1} \trans{\Vxi_i(\Vparam)} \\
=& - \frac{1}{2} tr\left(\Omega(\Vparam)^{-1} \dpartial{\Omega(\Vparam)}{\param}\right) 
 +  \Vxi_i(\Vparam) \Omega(\Vparam)^{-1} \trans{\dpartial{\Vmu(\Vparam,\VX_i)}{\param}} 
 + \frac{1}{2} \Vxi_i(\Vparam) \Omega(\Vparam)^{-1} \dpartial{\Omega(\Vparam)}{\param} \Omega(\Vparam)^{-1} \trans{\Vxi_i(\Vparam)} 
\end{align*}
Since \(\Esp\left[\Vxi_i(\Vparam)\right]=0\) and
\(\Esp\left[\trans{\Vxi_i(\Vparam)}\Vxi_i(\Vparam)\right] = \Omega(\Vparam)\), we obtain:
\begin{align*}
B &= \trans{\frac{\partial \Vmu(\Vparam,\VX_i)}{\partial \Vparam}} \Information(\Vparam)^{-1}
+ \frac{1}{2n} \sum_{j=1}^n \Esp\left[M_{ij}\right]\Information(\Vparam)^{-1} 
+ o_p(n^{-1}) 
\end{align*}
where \(M_{ij}\) is a matrix with elements:
\begin{align*}
(M_{ij})_{k,l} = \Vxi_{i,k}(\Vparam) \Vxi_j(\Vparam) \Omega(\Vparam)^{-1} \frac{\partial \Omega(\Vparam)}{\partial \param_l} \Omega(\Vparam)^{-1} \trans{\Vxi_j(\Vparam)}
\end{align*}
where \(\Vxi_{i,k}(\Vparam)\) denotes the residuals of the \(i\)-th
sample relative to the \(k\)-th endogenous variable. 

If \(i \neq j\), then \(\Vxi_{i,k}(\Vparam)\) is independent of
\(\Vxi_{j}(\Vparam)\) so \(\Esp\left[(M_{ij})_{k,l}\right]=0\). If
\(i=j\):
\begin{align*}
(M_{ij})_{k,l} = \Vxi_{i,k}(\Vparam)^3  A_{k,k} + 2 \Vxi_{i,k}(\Vparam)^2 \sum_{m'=1,m'\neq k}^M \Vxi_{i,m'} A_{m',k} 
+ \Vxi_{i,k}(\Vparam) \sum_{m'=1,m'\neq k}^M \sum_{m''=1,m''\neq k}^M \Vxi_{i,m'}(\Vparam) A_{m',m''} \Vxi_{i,m''}(\Vparam)
\end{align*}
So we only need to consider three cases: \(\Esp\left[z^3\right]\),
\(\Esp\left[z^2 x\right]\) and \(\Esp[x y z]\) where
(\(x\),\(y\),\(z\)) are jointly normally distributed with null
expectation and covariance \(\Omega\). In all cases the expectation is
null. Therefore \(\Esp\left[(M_{ij})_{k,l}\right]=0\) and:
\begin{align*}
B = \trans{\frac{\partial \Vmu(\Vparam,\VX_i)}{\partial \Vparam}} \Information(\Vparam)^{-1} + o_p(n^{-1}) 
\end{align*}

\begin{itemize}
\item \textbf{term C}: From lemma \ref{lemma:iid2}, \(C=o_p(n^{-1})\)

\item \textbf{term D}: \(\Esp\left[\trans{(\VparamHat - \Vparam)}(\VparamHat - \Vparam)\right] = \Sigma_{\VparamHat}\)
\end{itemize}

\bigskip

So using that \(O_p(||\VparamHat - \Vparam||^3)=o_p(n^{-1})\) we
obtain:
\begin{align*}
\Esp \left[\trans{\Vxi_i(\VparamHat)} \Vxi_i(\VparamHat) \right]
&= \Omega(\Vparam) - 2 \trans{\dpartial{\Vmu(\Vparam,\VX_i)}{\Vparam}} \Information(\Vparam)^{-1} \dpartial{\Vmu(\Vparam,\VX_i)}{\Vparam}
 + \trans{\dpartial{\Vmu(\Vparam,\VX_i)}{\Vparam}} \Sigma_{\VparamHat} \dpartial{\Vmu(\Vparam,\VX_i)}{\Vparam}
 + o_p(n^{-1}) \\
&= \Omega(\Vparam) - \trans{\dpartial{\Vmu(\Vparam,\VX_i)}{\Vparam}}  \Sigma_{\VparamHat} \dpartial{\Vmu(\Vparam,\VX_i)}{\Vparam} + o_p(n^{-1}) \\
\end{align*}

\subsection{Variance of the corrected residuals}
\label{sec:org390d5c7}

We define the corrected residuals as:
\begin{align}
\Vxi^c_i(\VparamHat) &= \Vxi_i(\VparamHat) \Omega(\Vparam)^{\half} \left(\Omega(\Vparam)^2-\Omega(\Vparam)^{\half}\Psi_i\Omega(\Vparam)^{\half}\right)^{-\half} \Omega(\Vparam)^{\half} \label{eq:adjResiduals} \\
&= \Vxi_i(\VparamHat) \Omega(\Vparam)^{\half} H_i^{-\half} \Omega(\Vparam)^{\half}  \notag
\end{align}
and note that to be well defined, \(\Omega(\Vparam)^2 -
 \Omega(\Vparam)^{\half} \Psi_i \Omega(\Vparam)^{\half}\) must be
 positive definite, i.e., the bias \(\Psi_i\) must be smaller (in
 terms of eigenvalues) than the initial estimate \(\Omega(\Vparam)\).

\bigskip

\uline{Example (standard linear model):} using that \(\Psi_i = \sigma^2
\VX_i \left(\trans{\VX} \VX\right)^{-1}\trans{\VX}_i\) and
\(\Omega(\Vparam)=\sigma^2\), formula \eqref{eq:adjResiduals} simplifies
into: \(\Vxi_i^c(\VparamHat) = \frac{ Y_{i} - \VX_{i} \hat{\beta}
}{1 - \VX_i \left(\trans{\VX} \VX\right)^{-1}\trans{\VX}_i}\). It is a
classical result that \(\Var\left[\Vxi_i^c(\VparamHat)\right] =
\sigma^2\). In fact, \(\Vxi_i^c(\VparamHat)/\sigma\) is the scaled
residual for observation \(i\) (\cite{galecki2013linear}, section
4.5.2).

\bigskip

We can check that the corrected residuals have mean 0:
\begin{align*}
\Esp\left[\Vxi^c_i(\VparamHat)\right] = \Esp\left[\Vxi_i(\VparamHat)\right] \Omega(\Vparam)^{\half} H_i^{-\half} \Omega(\Vparam)^{\half}=0
\end{align*}
and variance \(\Omega(\Vparam)\) up to a factor \(o_p(n^{-1})\):
\begin{align*}
&\Esp\left[\trans{\Vxi^c_i(\VparamHat)}\Vxi^c_i(\VparamHat)\right] \\
&=   \Omega(\Vparam)^{\half} H_i^{-\half} \Omega(\Vparam)^{\half} \Esp\left[\trans{\Vxi_i(\VparamHat)}\Vxi_i(\VparamHat)\right]  \Omega(\Vparam)^{\half} H_i^{-\half} \Omega(\Vparam)^{\half} \\
&=  \Omega(\Vparam)^{\half} H_i^{-\half} \Omega(\Vparam)^{\half} \left(\Omega(\Vparam)-\Psi_i+o_p(n^{-1}) \right)  \Omega(\Vparam)^{\half} H_i^{-\half} \Omega(\Vparam)^{\half}\\
&= \Omega(\Vparam)^{\half} H_i^{-\half} H_i H_i^{-\half} \Omega(\Vparam)^{\half} + o_p(n^{-1})  \\
&= \Omega(\Vparam) + o_p(n^{-1})  
\end{align*}

\subsection{Formula for the effective sample size}
\label{SM:effectiveSampleSize}
We can obtain a more explicit formula for the effective sample sizes
\(\Vn^{c}\). As shown in equation \eqref{main-eq:correctedN},
\(\Vn^{c}\) depends on
\(\dpartial{\Vxi_i(\VparamHat)}{\VY_i}=1-\dpartial{\Vmu(\VparamHat,\VX_i)}{\VY_i}\). Substituing
the expected to the observed information matrix in formula 2.2. of
\citet{wei1998generalized}, we obtain the following expression for the
generalized leverage in LVMs:
\begin{align}
\dpartial{\Vmu(\VparamHat,\VX_i)}{\VY_i} 
&= \trans{\dpartial{\Vmu(\VparamHat,\VX_i)}{\VparamHat}} \Information(\VparamHat)^{-1} \dpartial{\Score(\VparamHat|\VX_i)}{\VY_i} + o_p(n^{-\half}) \label{eq:leverage} \\
\text{where } \forall \param \in \Vparam \; \text{:} \; 
\dpartial{\Score(\param|\VX_i)}{\VY_{i}} &= \dpartial{\Vmu(\Vparam,\VX_i)}{\param} \Omega(\Vparam)^{-1} 
+ 2 \Vxi_i(\Vparam) \Omega(\Vparam)^{-1} \dpartial{\Omega(\Vparam)}{\param} \Omega(\Vparam)^{-1} \label{eq:dScoredY}
\end{align}

We will therefore modify algorithm 1 in two ways: first, we will
estimate \(\Vn^c\). We will estimate the corrected residuals and plug
them into equation \eqref{eq:dScoredY} because it involves
\(\Vxi_i(\Vparam)\). Second, when estimating the expected information
matrix, we will account for the fact that our effective sample size is
only \(\Vn^c\) by substituting, for each endogenous variable, the
effective sample size to \(n\) in first term of equation
\eqref{main-eq:Information}. More formally, as a substitute for \(n \;
tr(X)\) we introduce a new trace operator that for a \(T\) dimensional
squared matrix \(X\) is \(tr(X,\Vn^c) = \sum_{t=1}^T n^{c}_t
X_{t,t}\). The estimator for the expected information matrix becomes:
\begin{align}
\Information(\param,\param',\Vn^c) =& \frac{1}{2} tr\left(\Omega(\Vparam)^{-1} \dpartial{\Omega(\Vparam)}{\param} \Omega(\Vparam)^{-1} \dpartial{\Omega(\Vparam)}{\param'}, \Vn^c\right) + \sum_{i=1}^n \dpartial{\Vmu(\Vparam,\VX_i)}{\param} \Omega(\Vparam)^{-1} \trans{\dpartial{\Vmu(\Vparam,\VX_i)}{\param'}} 
\label{eq:InformationC}
\end{align}
This leads to the following algorithm:

\RestyleAlgo{boxruled}
\SetAlgoRefName{2}
\begin{algorithm}[H]
\textbf{Initialize:} 
\(\hat{\Omega}^{(0)} \leftarrow \hat{\Omega}(\Vparam)\) ;
 \(\hat{\Information}^{(0)} \leftarrow \Information(\VparamHat)\) ;
\(\dpartial{\hat{\Omega}^{(0)}}{\Vparam}=\dpartial{\Omega(\VparamHat)}{\Vparam}\)\\
\For{k=1 to \(\infty\)}{
(i) Compute for each subject \(i\) the bias
  \(\hat{\Psi}_i^{(k)}\) by plugging 
  \(\left(\hat{\Information}^{(k-1)}\right)^{-1}\) and 
  \(\dpartial{\Vmu(\VparamHat,\VX_i)}{\Vparam}\) into
  equation \eqref{main-eq:biasOmega}. \\ [2mm]
(ii) Compute the average bias \(\hat{\Psi}^{(k)} =
  \frac{1}{n}\sum_{i=1}^n \hat{\Psi}_i^{(k)}\). \\ [2mm]
(iii) Compute the corrected residuals \(\hat{\Vxi}_i^{(k)}\) plugging \(\hat{\Omega}^{(k-1)}\), \(\hat{\Psi}_i^{(k-1)}\) 
in equation \eqref{eq:adjResiduals}. \\ 
Compute the effective sample size \(\hat{\Vn}^{(k)}\) 
plugging \(\hat{\Omega}^{(k-1)}\), \(\hat{\Information}^{(k-1)}\), \(\hat{\Vxi}^{(k)}\), \(\dpartial{\mu(\VparamHat,\VX_i)}{\Vparam}\), and \(\dpartial{\hat{\Omega}^{(k-1)}}{\Vparam}\)
in equations \eqref{main-eq:correctedN}, \eqref{eq:leverage}, and \eqref{eq:dScoredY}. \\ [2mm]
(iv) Estimate \(\hat{\Omega}^{(k)}\) by \(\Omega(\VparamHat)+\hat{\Psi}^{(k)}\). \\ [2mm]
(v) Find \(\VparamHat_\Sigma^{(k)}\) satisfying \(\hat{\Omega}^{(k)} = \trans{\hat{\Lambda}} (1-\hat{B})^{-\intercal} \Sigma_\zeta\left(\VparamHat_\Sigma^{(k)}\right) (1-\hat{B})^{-1} \hat{\Lambda} + \Sigma_{\varepsilon} \left(\VparamHat_{\Sigma}^{(k)}\right) \) \\ [2mm]
(vi) Compute the corrected derivatives \(\dpartial{\hat{\Omega}^{(k)}}{\Vparam}\) using \(\VparamHat_\Sigma^{(k)}\) \\ [2mm]
(vii) Compute \(\hat{\Information}^{k}\) by plugging \(\hat{\Omega}^{(k)}\), \(\hat{\Vn}^{(k)}\), \(\dpartial{\mu(\VparamHat,\VX_i)}{\Vparam}\), and \(\dpartial{\hat{\Omega}^{(k)}}{\Vparam}\) into equation \eqref{eq:InformationC}
}
\caption{}
\end{algorithm}

\clearpage

\section{Standard linear model}
\label{SM:correctLM}
We consider an iid sample \((Y_i,\VX_i)_{i\in\{1,\ldots,n\}}\) of two
random variables \(Y,\VX\) and the following linear model:
\begin{align*}
Y_i = \VX_i \beta + \varepsilon_i \text{, where } \varepsilon_i \sim \Gaus\left(0,\sigma^2\right)
\end{align*}
Denoting \(\Vparam = (\beta,\sigma^2)\), the ML estimator for \(\Vparam\) is
is:
\begin{align*}
\hat{\beta} &= \left(\trans{\VX}\VX\right)^{-1} \trans{\VX} Y \\
\hat{\sigma}^2 &= \frac{1}{n} \sum_{i=1}^n \left(Y_i - \VX_i \hat{\beta}\right)^2
\end{align*}

\bigskip

We can also write the likelihood, the score, the Hessian, and the expected information matrix of this
model:
\begin{align*}
l(\Vparam) &= \sum_{i=1}^n \left[ - \frac{1}{2} \log(2\pi) - \frac{1}{2} \log(\sigma^2) - \frac{1}{2\sigma^2}(Y_i-\VX_i\beta)^2 \right] \\
\Score(\Vparam) &= \begin{bmatrix} 
\sum_{i=1}^n \frac{\trans{\VX_i}}{\sigma^2} (Y_i - \VX_i \beta) \\
- \sum_{i=1}^n \left( \frac{1}{2\sigma^2} - \frac{1}{2\sigma^4} (Y_i - \VX_i \beta)^2 \right) \\
\end{bmatrix} \\
\Hessian(\Vparam) &= \begin{bmatrix} 
-\sum_{i=1}^n \frac{\trans{\VX_i} \VX_i}{\sigma^2} & -\sum_{i=1}^n \frac{\trans{\VX_i}}{\sigma^4}(Y_i-\VX_i\beta) \\
-\sum_{i=1}^n \frac{\trans{\VX_i}}{\sigma^4}(Y_i-\VX_i\beta) & \sum_{i=1}^n \left( \frac{1}{2\sigma^4} - \frac{1}{\sigma^6} (Y_i - \VX_i \beta)^2 \right) \\
\end{bmatrix} \\
\Information(\Vparam) &= \begin{bmatrix} 
\frac{\trans{\VX} \VX}{\sigma^2} & 0\\
0 &  \frac{n}{2\sigma^4} \\
\end{bmatrix} \\
\end{align*}

\subsection{Correction of the bias of the ML-estimator}
\label{SM:LM-correctBias}
We use equation \eqref{main-eq:biasOmega} to estimate the bias of
\(\hat{\sigma}^2\):
\begin{align*}
\Psi_i &= 
\begin{bmatrix} \VX_i & 0 \end{bmatrix} 
\begin{bmatrix} \sigma^2 \left(\trans{\VX}\VX\right)^{-1} & 0 \\ 0 & \frac{2 \sigma^4}{n} \end{bmatrix}
\begin{bmatrix} \trans{\VX_i} \\ 0 \end{bmatrix} 
= \VX_i \left(\trans{\VX}\VX\right)^{-1} \trans{\VX_i}\sigma^2
\end{align*}
So we have:
\begin{align*}
\Psi &= \frac{1}{n} \sum_{i=1}^n \Psi_i = \frac{p}{n}\sigma^2 \\
\hat{\Psi} &= \frac{p}{n}\hat{\sigma}^2
\end{align*}
and therefore:
\begin{align*}
\left(\hat{\sigma}^{c}\right)^2 &= \left(1+\frac{p}{n}\right) \hat{\sigma}^2 
\end{align*}
We can compute the expectation of \(\left(\hat{\sigma}^{c}\right)^2\):
\begin{align*}
\Esp\left[\left(\hat{\sigma}^{c}\right)^2\right] &= \left(1-\frac{p^2}{n^2}\right) \sigma^2
\end{align*}
and see that the bias is now \(\frac{p^2}{n^2}\) instead of
\(\frac{p}{n}\). To get an unbiased estimate of \(\hat{\sigma}^c\), we
would need to use Algorithm 1, i.e., iterate between:
\begin{itemize}
\item estimating \(\hat{\Psi}^{(k)} = \hat{\sigma}^2 \sum_{\kappa=1}^{k} \frac{p^\kappa}{n^\kappa}\)
\item estimating \(\left(\hat{\sigma}^{c}\right)^2 = \hat{\sigma}^2 + \hat{\Psi}^{(k)} = \hat{\sigma}^2 \sum_{\kappa=0}^{k} \frac{p^\kappa}{n^\kappa}\)
\end{itemize}
where \(k\) is the iteration number. Since 
\begin{align*}
\sum_{\kappa=0}^{k} \left(\frac{p}{n}\right)^{\kappa} \xrightarrow[k\rightarrow \infty]{} \frac{1}{1-\frac{p}{n}} = \frac{n}{n-p}
\end{align*}
and assuming \(n>p\), we will get the usual unbiased estimator for \(\sigma^2\):
\begin{align*}
\left(\hat{\sigma}^c\right)^2 &\xrightarrow[k\rightarrow \infty]{} \frac{1}{n-p} \sum_{i=1}^n \left(Y_i - \VX_i \hat{\beta}\right)^2
\end{align*}
\subsection{Satterthwaite approximation}
\label{SM:LM-correctSatterthwaite}
The Satterthwaite correction aims to estimate \(df\) such that
\(\frac{\hat{\beta}}{\hat{\sigma}_{\hat{\beta}}}\) is approximately
distributed as a Student's \(t\) distribution with variance 1 and
degrees of freedom \(df\). As shown in equation
\eqref{main-eq:dfSatterthwaite}, we can estimate \(df\) with the help of
an estimate of \(\Var\left[\sigma^2_{\hat{\beta}}\right]\). Therefore
we need to use only a delta method and to compute the first derivative of
the expected information matrix. From
\(\Esp\left[Y_i-\VX_i\beta\right]=0\) and
\(\Esp\left[(Y_i-\VX_i\beta)^2\right]=\sigma^2\), we obtain that the
expected information matrix is:
\begin{align}
\Information(\Vparam) &= \begin{bmatrix} 
\frac{\trans{\VX}\VX}{\sigma^2} & 0 \\
0 & \frac{n}{2\sigma^4} \\
\end{bmatrix} \notag
\end{align}
The first derivative of the information matrix is:
\begin{align*}
\nabla_{\Vparam} \Information(\Vparam) &= \left( \nabla_{\beta} \Information(\Vparam), \nabla_{\sigma^2} \Information(\Vparam) \right) \\
&= \left(
\begin{bmatrix} 
0 & 0 \\
0 & 0 \\
\end{bmatrix},
\begin{bmatrix} 
-\frac{\trans{\VX}\VX}{\sigma^4} & 0 \\
0 & -\frac{n}{\sigma^6} \\
\end{bmatrix}
\right)
\end{align*}
Since \(\Sigma_{\VparamHat} = \Information(\Vparam)^{-1}\):
\begin{align}
\nabla_{\Vparam} \Sigma_{\VparamHat} &= \left( \nabla_{\beta} \Sigma(\Vparam), \nabla_{\sigma^2} \Sigma(\Vparam) \right) \notag \\
&= \left(- \Information(\Vparam)^{-1} \nabla_{\beta} \Information(\Vparam) \Information(\Vparam)^{-1},
         - \Information(\Vparam)^{-1} \nabla_{\sigma^2} \Information(\Vparam) \Information(\Vparam)^{-1} \right) \notag \\
&= \left(
\begin{bmatrix} 
0 & 0 \\
0 & 0 \\
\end{bmatrix},
\begin{bmatrix} 
\sigma^2 \left(\trans{\VX}\VX\right)^{-1} & 0 \\
0 & \frac{2\sigma^4}{n} \\
\end{bmatrix}
\begin{bmatrix} 
\frac{\trans{\VX}\VX}{\sigma^4} & 0 \\
0 & \frac{n}{\sigma^6} \\
\end{bmatrix}
\begin{bmatrix} 
\sigma^2 \left(\trans{\VX}\VX\right)^{-1} & 0 \\
0 & \frac{2\sigma^4}{n} \\
\end{bmatrix} 
\right) \notag \\
&= \left(
\begin{bmatrix} 
0 & 0 \\
0 & 0 \\
\end{bmatrix},
\begin{bmatrix} 
\left(\trans{\VX}\VX\right)^{-1} & 0 \\
0 & \frac{4\sigma^2}{n} \\
\end{bmatrix}
\right) \notag
\end{align}
Considering the \(j\)-th coefficient \(\beta_j\) and denoting \(c_j\)
the vector with a one at the \(j\)-th position and zeros otherwise, we
have:
\begin{align*}
\Var\left[\sigma^2_{\hat{\beta}_j}\right] &= \nabla_{\Vparam} \sigma^2_{\hat{\beta}_j}(\Vparam) \Sigma(\Vparam) \nabla_{\Vparam} \sigma^2_{\hat{\beta}_j}(\Vparam) \\
&= \begin{bmatrix} 
0 & c_j\left(\trans{\VX}\VX\right)^{-1}\trans{c_j} \\
\end{bmatrix}
\begin{bmatrix} 
\sigma^2 \left(\trans{\VX}\VX\right)^{-1} & 0 \\
0 & \frac{2\sigma^4}{n} \\
\end{bmatrix}
\begin{bmatrix} 
0 \\ c_j\left(\trans{\VX}\VX\right)^{-1}\trans{c_j} \\ 
\end{bmatrix} \\
&= \frac{2}{n}  \left(\sigma^2 c_j\left(\trans{\VX}\VX\right)^{-1} \trans{c_j} \right)^2 =  2 \frac{\left(\sigma_{\hat{\beta}_j}\right)^4}{n}
\end{align*}
and
\begin{align*}
\widehat{df}(\beta_j) = 2 \frac{\left(\sigma_{\hat{\beta}_j}\right)^4}{2 \frac{\left(\sigma_{\hat{\beta}_j}\right)^4}{n}} = n
\end{align*}
Therefore, direct application of the Satterthwaite approximation gives
\(n\) for the degrees of freedom of the Wald statistic.

\subsection{Satterthwaite approximation using the effective sample size}
\label{SM:LM-correctDF}
Denoting \(p\) as the rank of \(X\), i.e., the number of \(\beta\)
coefficients, the effective sample size \(\hat{n}^c\) can be computed
using equation \eqref{main-eq:correctedN} :
\begin{align*}
\hat{n}^c = n - \sum_{i=1}^n \VX_i \dpartial{\beta}{Y_i} 
= n - \sum_{i=1}^n \VX_i \left(\trans{\VX}\VX\right)^{-1} \trans{\VX_i} = n - p
\end{align*}
In this specific model, \(\hat{n}^c\) can be obtained without
iteration because it does not depend on \(\sigma^2\). So the corrected
information matrix becomes:
\begin{align}
\Information(\Vparam) &= \begin{bmatrix} 
\frac{\trans{\VX}\VX}{\sigma^2} & 0 \\
0 & \frac{n-p}{2\sigma^4} \\
\end{bmatrix} \notag
\end{align}
and the degrees of freedom becomes:
\begin{align*}
\widehat{df}(\beta_j) &= \hat{n}^c = n - p
\end{align*}

\clearpage

\section{Mean-variance model}
\label{sec:orgf5fba03}

We consider a LVM where no parameter appears both in the conditional
mean and in the conditional variance. So for a given parameter
\(\param \in \set_{\Vparam}\),
\(\dpartial{\Vmu(\Vparam,\VX_i)}{\param}\) and
\(\dpartial{\Omega(\Vparam)}{\param}\) cannot be simultaneously non-0.
The information matrix is then block diagonal with one block relative
to the mean parameters (\(\Vparam_\mu\)) and one block relative to the
variance parameters (\(\Vparam_\Sigma\)). We also denote by \(p_\mu\)
the number of mean parameters.

\subsection{Convergence of Algorithm 1}
\label{SM:algo1}
Injecting the right
hand side of equation \eqref{main-eq:Information} in \eqref{main-eq:biasOmega}
gives:
\begin{align*}
\Psi &= \frac{1}{n} \sum_{i=1}^n \trans{\dpartial{\Vmu(\Vparam,\VX_i)}{\Vparam_\mu}}
 \left(\sum_{i=1}^n \dpartial{\Vmu(\Vparam,\VX_i)}{\Vparam_\mu} (\Omega(\Vparam)+\Psi)^{-1} \trans{\dpartial{\Vmu(\Vparam,\VX_i)}{\Vparam_\mu}} \right)^{-1} 
\dpartial{\Vmu(\Vparam,\VX_i)}{\Vparam_\mu}
\end{align*}
This is the fixed point equation that we are solving with
Algorithm 1. 

\bigskip

\textbf{Monotonicity}: we use the previous equation to show that Algorithm 1
gives a sequence of increasing \(\Psi^{(k)}\) in the sense that for
any iteration \(k\geq 2\), \(\Delta \hat{\Psi}^{(k)} =
\hat{\Psi}^{(k)}-\hat{\Psi}^{(k-1)}\) is semi-definite
positive. Indeed, if \(\Delta \hat{\Psi}^{(k-1)}\) is semi-definite
positive, then \(\hat{\Omega}^{(k-1)}-\hat{\Omega}^{(k-2)}\) is
semi-definite positive and
\(\left(\hat{\Omega}^{(k-1)}\right)^{-1}-\left(\hat{\Omega}^{(k-2)}\right)^{-1}\)
is semi-definite negative. Applying similar arguments to
\(\left(\sum_{i=1}^n \dpartial{\Vmu(\VparamHat,\VX_i)}{\Vparam_\mu}
\left(\hat{\Omega}^{(k-1)}\right)^{-1}
\trans{\dpartial{\Vmu(\VparamHat,\VX_i)}{\Vparam_\mu}} \right)^{-1}\)
leads to \(\Delta \hat{\Psi}^{(k)}\) being semi-definite positive.

\bigskip

\textbf{Convergence}: denoting \(Z_i = \dpartial{\Vmu(\Vparam,\VX_i)}{\Vparam_\mu}
(\Omega(\Vparam)+\Psi)^{-\half}\) and \(Z = \sum_{i=1}^n \trans{Z_i}
\left(\sum_{i=1}^n Z_i \trans{Z_i} \right)^{-1} Z_i\) we get:
\begin{align*}
\Psi = \frac{1}{n} (\Omega(\Vparam)+\Psi)^{-\half} Z (\Omega(\Vparam)+\Psi)^{-\half} 
\end{align*}
Using that the trace is linear and is invariant under cyclic
permutation, we obtain that \(tr(Z)=p_\mu\). \(Z\), being semi-definite
positive, has eigenvalues that are bounded by \(p_\mu\). Therefore, at iteration
\(k>2\), one can show by induction that we have:
\begin{align*}
||\hat{\Psi}^{(k)}|| \leq \frac{p_\mu}{n} ||\hat{\Omega}^{(k)}|| \leq \frac{p_\mu}{n}
\left(||\Omega(\VparamHat)|| + || \hat{\Psi}^{(k-1)} || \right) \leq
\frac{p_\mu}{n} \left(1+\sum_{\kappa=1}^{k-1}
\frac{p_{\mu}^{\kappa}}{n^{\kappa}}\right) ||\Omega(\VparamHat)||
\end{align*}
where \(||.||\) denotes the spectral norm. Therefore, a sufficient condition for
\(\hat{\Psi}^{(k)}\) to be bounded above (therefore, for Algorithm 1 to
converge) is that \(p_\mu < n\).

\subsection{Effective sample size}
\label{sec:org0420a1b}

We have that \(\trans{\dpartial{\Vmu(\VparamHat,\VX_i)}{\VparamHat}}
\Information(\VparamHat)^{-1}\) is block diagonal with only the block
relative to \(\Vparam_\mu\) is non-0. Therefore, from equations
\eqref{eq:leverage} and \eqref{eq:dScoredY}, we get:
\begin{align}
\dpartial{\Vmu(\VparamHat,\VX_i)}{\VY_i} =
 \trans{\dpartial{\Vmu(\VparamHat,\VX_i)}{\VparamHat}} \Information(\VparamHat)^{-1}  \dpartial{\Vmu(\VparamHat,\VX_i)}{\Vparam} \Omega(\VparamHat)^{-1}  + o_p(n^{-\half})
\end{align}
(the second term, which includes
\(\dpartial{\Omega(\Vparam)}{\Vparam}\), becomes 0 when multiplied by
\(\trans{\dpartial{\Vmu(\VparamHat,\VX_i)}{\VparamHat}}
\Information(\VparamHat)^{-1}\)).

\clearpage

\section{Multivariate Wald tests}
\label{sec:orge776a39}
\subsection{Definition}
\label{SM:multivariateWaldDef}
We now consider testing several null hypotheses simultaneously. For
example, given three different parameters (e.g., the first three
parameters denoted \(\param_1\), \(\param_2\), and \(\param_3\)), we
wish to test the global null hypothesis \(\left(\Hypothesis_0\right)\)
\(\param_1=0\), \(\param_2=0\) and \(\param_3=0\). To do so, we define
a contrast matrix \(C\) encoding a hypothesis in each line. In the
previous example, \(C\) is a \(3\) by \(p\) matrix full of zeros
except at the first three elements of the diagonal which contain
one. The global null hypothesis can be tested using following Wald
statistic:
\begin{align*}
\mathcal{F}_{C \Vparam} = \frac{1}{Q} \trans{(C \Vparam)} (C \Sigma_{\VparamHat} \trans{C})^{-1} (C \Vparam)
\end{align*}
where \(Q\) is the rank of \(C\). As in the univariate case, by
plugging in \(\hat{\Sigma}_{\VparamHat}\) and using the asymptotic
normality of \(\paramHat\), one obtains that \(\mathcal{F}_{C
\Vparam}\) is asymptotically chi-squared distributed with \(Q\) degrees
of freedom.

\subsection{Degrees of freedom}
\label{SM:multivariateWaldDf}
In a finite sample, ignoring the effect of the plug-in estimate will
lead to an incorrect control of the type 1 error rates. A possible
remedy is to use a Fisher distribution, instead of a chi-squared
distribution, to account for the variability of
\(\hat{\Sigma}_{\VparamHat}\) when modeling the distribution of
\(\hat{\mathcal{F}}_{C \Vparam}\). As proposed in
\citet{hrong1996approximate}, the Satterthwaite approximation can be used
to estimate the degrees of freedom of the distribution. Here, we will
only consider the one-moment approximation, which approximates the
distribution of \(\mathcal{F}_{C\Vparam}\) by a Fisher distribution
with \(Q\) and \(\nu\) degrees of freedom. \(Q\) is known, but \(\nu\)
needs to be estimated.

\bigskip

 Using an eigenvalue decomposition, one can show that \(Q
\mathcal{F}_{C\Vparam}\) can be written as a sum of \(Q\) independent
and \(t\)-distributed variables squared:
\begin{align*}
Q \mathcal{F}_{C\Vparam} = \sum_{q=1}^Q  \frac{(P C \Vparam)^2_q}{\kappa_q}
\end{align*}
where \((\kappa_q)_{q \in \{1,\ldots,Q\}}\) are the eigenvalues of \(Q
\mathcal{F}_{C\Vparam}\), \(P\) a corresponding matrix of eigenvectors,
and \((\nu_q)_{q\in \{1,\ldots,Q\}}\) the degrees of freedom of the
Student's \(t\)-distributions. This can be seen as defining a new set
of contrasts \(\left(\frac{(P C)_q}{\sqrt{\kappa_q}}\right)_{q \in
\{1,\ldots,Q\}}\) and equation \eqref{main-eq:dfSatterthwaite} can be
used to estimate \((\nu_q)_{q \in \{1,\ldots,Q\}}\). Finally, because,
\(\Esp\left[\mathcal{F}_{C\Vparam}\right] = \frac{\nu}{\nu-2}\) and
\(\Esp\left[\frac{(P C \param)^2_q}{\kappa_q}\right] =
\frac{\nu_q}{\nu_q-2}\), we get:
\begin{align*}
\nu = \frac{2 \sum_{q=1}^Q \frac{\nu_q}{\nu_q-2}}{\sum_{q=1}^Q \frac{\nu_q}{\nu_q-2} - Q}
\end{align*}

\clearpage

\section{Simulation study}
\label{SM:figSimulation}
\subsection{Convergence issues}
\label{SM:cvIssues}
We met three types of issues when performing the
simulations in small samples:
\begin{itemize}
\item the optimizer used to estimate the LVM parameters did not always
converge correctly: non-zero gradient at the end of the procedure,
singular information matrix, or standard errors extremely large
(>1000).  For n=20, this concerned less than 1\% of the simulations
in study A, around 10\% of the simulations in study B for n=20,
and around 16\% of the simulations for study C. For n=30, it
affected less than 3\% of simulations and very few simulations
for n=50.
\item In study C and n=20, \(\Omega(\Vparam)^2 - \Omega(\Vparam)^{\half}
  \Psi_i \Omega(\Vparam)^{\half}\) was not positive definite in 7
simulations (<1\%), i.e., the corrected residuals were not well
defined.
\item In study C and n=20, Algorithm 2 failed to converge in 11
simulations (<1\%). Otherwise, convergence was reached in few
iterations, e.g., in study C an average of 11 iterations at n=20
and 4 iterations at n=500.
\end{itemize}
Simulations with the first type of issue were discarded. When the
second type of issue was met, Algorithm 1 was used instead of
Algorithm 2. Simulations with the second or third type of issue were
analyzed normally.

\clearpage

\subsection{Bias of the ML-estimator in small samples}
\label{SM:tableBiasML}
\subsubsection{Study A}
\label{sec:orgd4e44f6}
\begin{table}[!h]
\begin{tabular}{c|c|cccc} 
 \multirow{2}{*}{sample size} & \multirow{2}{*}{type} & \multicolumn{2}{c}{ML} & \multicolumn{2}{c}{ML-corrected} \\ 
 & &  mean (sd) & median [Q1;Q3] & mean (sd) & median [Q1;Q3] \\ \hline 
 20 & $\Sigma_{\varepsilon}$ & -0.050 (0.218) & -0.067 [-0.204;0.086] & <0.001 (0.229) & -0.018 [-0.162;0.143] \\ 
  & $\Sigma_{\zeta}$ & -0.182 (0.394) & -0.227 [-0.469;0.058] & 0.002 (0.463) & -0.052 [-0.335;0.282] \\ 
 30 & $\Sigma_{\varepsilon}$ & -0.034 (0.180) & -0.045 [-0.161;0.080] & -0.001 (0.187) & -0.012 [-0.132;0.117] \\ 
  & $\Sigma_{\zeta}$ & -0.128 (0.330) & -0.157 [-0.362;0.075] & -0.006 (0.366) & -0.039 [-0.266;0.219] \\ 
 50 & $\Sigma_{\varepsilon}$ & -0.020 (0.140) & -0.027 [-0.118;0.071] & <0.001 (0.143) & -0.007 [-0.100;0.092] \\ 
  & $\Sigma_{\zeta}$ & -0.071 (0.264) & -0.089 [-0.260;0.098] & 0.002 (0.281) & -0.017 [-0.198;0.182] \\ 
 100 & $\Sigma_{\varepsilon}$ & -0.011 (0.099) & -0.014 [-0.080;0.054] & -0.001 (0.100) & -0.004 [-0.071;0.065] \\ 
  & $\Sigma_{\zeta}$ & -0.037 (0.189) & -0.046 [-0.169;0.087] & <0.001 (0.194) & -0.010 [-0.136;0.127] \\ 
 \end{tabular} 
\caption{Average bias (standard deviation) and median bias (first and third quantile) of the residual variance (\(\Sigma_{\varepsilon}\))
 and the variance of the latent variable \(\Sigma_{\zeta}\). 
In small samples, the bias for the other parameters was at least one order of magnitude smaller than for \(\Sigma_{\varepsilon}\). 
The symbol <0.001 indicates an absolute value smaller than 0.001.}
\label{tab:biasMM}
\end{table}

\subsubsection{Study B}
\label{sec:org8ad5385}
\begin{table}[!h]
\begin{tabular}{c|c|cccc} 
 \multirow{2}{*}{sample size} & \multirow{2}{*}{type} & \multicolumn{2}{c}{ML} & \multicolumn{2}{c}{ML-corrected} \\ 
 & &  mean (sd) & median [Q1;Q3] & mean (sd) & median [Q1;Q3] \\ \hline 
 20 & $\Sigma_{\varepsilon}$ & -0.125 (0.411) & -0.161 [-0.412;0.118] & -0.029 (0.454) & -0.071 [-0.346;0.238] \\ 
  & $\Sigma_{\zeta}$ & -0.150 (0.494) & -0.230 [-0.514;0.124] & 0.018 (0.585) & -0.075 [-0.411;0.343] \\ 
 30 & $\Sigma_{\varepsilon}$ & -0.085 (0.349) & -0.107 [-0.321;0.127] & -0.021 (0.372) & -0.046 [-0.273;0.204] \\ 
  & $\Sigma_{\zeta}$ & -0.096 (0.415) & -0.148 [-0.399;0.146] & 0.017 (0.464) & -0.041 [-0.320;0.287] \\ 
 50 & $\Sigma_{\varepsilon}$ & -0.052 (0.270) & -0.065 [-0.231;0.115] & -0.013 (0.280) & -0.027 [-0.200;0.161] \\ 
  & $\Sigma_{\zeta}$ & -0.055 (0.326) & -0.087 [-0.288;0.146] & 0.014 (0.348) & -0.020 [-0.235;0.228] \\ 
 100 & $\Sigma_{\varepsilon}$ & -0.026 (0.190) & -0.032 [-0.152;0.095] & -0.006 (0.194) & -0.013 [-0.135;0.117] \\ 
  & $\Sigma_{\zeta}$ & -0.026 (0.234) & -0.046 [-0.190;0.119] & 0.008 (0.242) & -0.012 [-0.161;0.158] \\ 
 \end{tabular} 
\caption{Average bias (standard deviation) and median bias (first and third quantile) of the residual variances (\(\Sigma_{\varepsilon}\))
 and the variance parameter of the latent variable \(\Sigma_{\zeta}\).
In small samples, the bias for the other parameters was at least one order of magnitude smaller than for \(\Sigma_{\varepsilon}\).
The symbol <0.001 indicates an absolute value smaller than 0.001.}
\label{tab:biasFactor}
\end{table}

\subsubsection{Study C}
\label{sec:orgb13a52f}
\begin{table}[!h]
\begin{tabular}{c|c|cccc} 
 \multirow{2}{*}{sample size} & \multirow{2}{*}{type} & \multicolumn{2}{c}{ML} & \multicolumn{2}{c}{ML-corrected} \\ 
 & &  mean (sd) & median [Q1;Q3] & mean (sd) & median [Q1;Q3] \\ \hline 
 20 & $\Sigma_{\varepsilon}$ & -0.110 (0.402) & -0.148 [-0.394;0.131] & -0.034 (0.437) & -0.076 [-0.342;0.227] \\ 
  & $\Sigma_{\zeta}$ & -0.142 (0.513) & -0.231 [-0.522;0.145] & -0.004 (0.584) & -0.104 [-0.438;0.324] \\ 
  & $\Sigma_{\varepsilon,\varepsilon'}$ & 0.015 (0.359) & 0.004 [-0.221;0.237] & 0.017 (0.392) & 0.005 [-0.241;0.259] \\ 
 30 & $\Sigma_{\varepsilon}$ & -0.075 (0.335) & -0.099 [-0.305;0.132] & -0.024 (0.353) & -0.050 [-0.267;0.193] \\ 
  & $\Sigma_{\zeta}$ & -0.094 (0.428) & -0.151 [-0.405;0.152] & <0.001 (0.467) & -0.061 [-0.340;0.272] \\ 
  & $\Sigma_{\varepsilon,\varepsilon'}$ & 0.001 (0.296) & -0.006 [-0.198;0.190] & 0.002 (0.313) & -0.005 [-0.209;0.202] \\ 
 50 & $\Sigma_{\varepsilon}$ & -0.046 (0.259) & -0.060 [-0.224;0.117] & -0.016 (0.267) & -0.030 [-0.198;0.153] \\ 
  & $\Sigma_{\zeta}$ & -0.052 (0.336) & -0.088 [-0.292;0.152] & 0.005 (0.353) & -0.031 [-0.248;0.220] \\ 
  & $\Sigma_{\varepsilon,\varepsilon'}$ & -0.003 (0.223) & -0.007 [-0.152;0.143] & -0.002 (0.231) & -0.007 [-0.156;0.149] \\ 
 100 & $\Sigma_{\varepsilon}$ & -0.023 (0.181) & -0.031 [-0.147;0.093] & -0.008 (0.184) & -0.016 [-0.134;0.110] \\ 
  & $\Sigma_{\zeta}$ & -0.026 (0.237) & -0.043 [-0.195;0.122] & 0.003 (0.243) & -0.015 [-0.170;0.154] \\ 
  & $\Sigma_{\varepsilon,\varepsilon'}$ & -0.001 (0.155) & -0.001 [-0.106;0.102] & -0.001 (0.158) & -0.001 [-0.107;0.104] \\ 
 \end{tabular} 
\caption{Average bias (standard deviation) and median bias (first and third quantile) of the residual variances (\(\Sigma_{\varepsilon}\)), 
of the covariance parameter (\(\Sigma_{\varepsilon,\varepsilon'}\)),  
and the variance parameters for the latent variable \(\Sigma_{\zeta}\).
In small samples, the bias for the other parameters was at least one order of magnitude smaller than for \(\Sigma_{\varepsilon}\).
The symbol <0.001 indicates an absolute value smaller than 0.001.}
\label{tab:biasLVM}
\end{table}

\clearpage

\subsection{Validity of the Student's \(t\)-distribution for modeling the distribution of the Wald statistic}
\label{SM:validityStudent}
We fitted, using maximum likelihood, a non-standardized Student's
\(t\)-distribution (i.e., Student's \(t\)-distribution with a
non-centrality parameter, a dispersion parameter, and a parameter for
the degrees of freedom) to the Wald statistics obtained from the
simulations with \(n=20\). We tested the adequacy of the resulting
distribution to the empirical distribution using a Kolmogorov Smirnov
test. For parameters where the Student's \(t\)-distribution was
appropriate, we compared the estimated dispersion to 1 (expected
dispersion of a Wald statistic) and the estimated degrees of freedom
to the average degrees of freedom estimated over the simulations
(called Satterthwaite degrees of freedom). The results are reported in
\autoref{tab:validation}.

\bigskip

In study A, we see that the estimated dispersion was very close to
1 and the estimated degrees of freedom was close to the Satterthwaite
degrees of freedom. The Student's \(t\)-distribution also appeared to
give a satisfactory fit. This confirms the validity of our correction
for random intercept models. In study B and C, the Student's
\(t\)-distribution was also satisfactory for all but two parameters
(\(b_1\) and \(\sigma_{12}\)). Visual inspection of the empirical
distribution of the Wald statistics for these parameters revealed a
slight skweness. The estimated dispersion appeared to vary between 1
and 1.15, meaning that the estimated standard error of some parameters
was biased. This is probably due to the fact the our small sample
correction did not entirely correct the bias of the estimator of the
variance parameters. Finally the average Satterthwaite degrees of
freedom appeared to differ sometimes substantially from the ones
estimated with the empirical distribution. This suggests that the
correction using the Satterthwaite approximation is not always
reliable in small samples. These elements may explain the imperfect
control of the type 1 error when using the proposed correction.

\begin{table}[ht]
\centering
\begin{tabular}{lllllllll}
  \toprule
  &&\multicolumn{2}{c}{Empirical Student} & \multicolumn{2}{c}{Modeled Student} & Expected
& \multicolumn{2}{c}{KS-test} \\ \cmidrule(lr){3-4} \cmidrule(lr){5-6} \cmidrule(lr){8-9} 
 Study & parameter & dispersion & df & dispersion & df & type 1 error & statistic & p-value  \\
 \midrule
 A & $\nu_2$ & 1.006 & 40.8 & 1 & 36.7 & 0.051 & 0.005 & 0.659 \\ 
   & $\gamma_1$ & 1.004 & 17.9 & 1 & 18.3 & 0.051 & 0.004 & 0.961 \\ 
   [4mm] B & $\nu_2$ & 0.998 & 39.2 & 1 & 9.1 & 0.029 & 0.006 & 0.513 \\ 
   & $\lambda_4$ & 1.067 & 117.2 & 1 & 5.9 & 0.023 & 0.004 & 0.885 \\ 
   & $\gamma_1$ & 1.088 & 311.8 & 1 & 10.5 & 0.043 & 0.006 & 0.455 \\ 
   & $k_1$ & 1.063 & 14.8 & 1 & 17.1 & 0.066 & 0.005 & 0.792 \\ 
   [4mm] C & $\nu_2$ & 1.026 & 45.5 & 1 & 10.1 & 0.035 & 0.006 & 0.536 \\ 
   & $k_1$ & 1.106 & 12.6 & 1 & 17.1 & 0.080 & 0.004 & 0.936 \\ 
   & $\lambda_4$ & 1.080 & 145.9 & 1 & 6.0 & 0.025 & 0.008 & 0.237 \\ 
   & $\gamma_1$ & 1.152 & 49.6 & 1 & 10.1 & 0.059 & 0.009 & 0.135 \\ 
   & $b_1$ & 1.127 & 783296.7 & 1 & 3.6 & 0.010 & 0.017 & <0.001 \\ 
   & $\sigma_{12}$ & 1.074 & 934003.5 & 1 & 7.8 & 0.031 & 0.038 & <0.001 \\ 
   \bottomrule
\end{tabular}
\caption{Comparison between the empirical distribution of the Wald statistic vs. the modeled distribution (after correction) for n=20.  Empirical Student: standard error and degrees of freedom of a non-standardized Student's \(t\)-distribution fitted to the empirical distribution. Modeled Student: average estimated degrees of freedom over the simulations. Expected type 1 error: rejection rate under the empirical Student when using the 2.5 and 97.5\% quantile of the modeled Student. Kolmogorov Smirnov test: comparison between the empirical cumluative distribution function (cdf) and the cdf of the empirical Student.} 
\label{tab:validation}
\end{table}

\clearpage

\subsection{Type 1 error of (uncorrected) Wald tests obtained with ML, GLS, WLS, or IV estimators}
\label{SM:comparison}
\begin{table}[ht]
\centering
\begin{tabular}{llrrrrrr}
  \toprule
parameter & n & ML & robust ML & GLS & WLS & IV (lava) & IV (lavaan) \\ 
  \midrule
$\alpha$ & 20 & 0.088 & 0.094 & 0.075 & NA & 0.108 & 0.087 \\ 
  $\nu_2$ &  & 0.082 & 0.093 & 0.067 & NA & 0.101 & 0.081 \\ 
  $\nu_3$ &  & 0.062 & 0.068 & 0.053 & NA & 0.075 & 0.070 \\ 
  $\nu_4$ &  & 0.073 & 0.074 & 0.057 & NA & 0.077 & 0.073 \\ 
  $k_1$ &  & 0.115 & 0.108 & 0.097 & NA & 0.088 & 0.081 \\ 
  $\gamma_1$ &  & 0.096 & 0.089 & 0.081 & NA & 0.094 & 0.085 \\ 
  $\gamma_2$ &  & 0.107 & 0.114 & 0.135 & NA & 0.126 & 0.088 \\ 
  $\lambda_2$ &  & 0.089 & 0.094 & 0.112 & NA & 0.161 & 0.128 \\ 
  $\lambda_3$ &  & 0.086 & 0.093 & 0.114 & NA & 0.180 & 0.143 \\ 
  $\lambda_4$ &  & 0.075 & 0.086 & 0.088 & NA & 0.101 & 0.067 \\ 
   [4mm]$\alpha$ & 50 & 0.062 & 0.064 & 0.061 & 0.046 & 0.068 & 0.062 \\ 
  $\nu_2$ &  & 0.058 & 0.062 & 0.055 & 0.092 & 0.063 & 0.057 \\ 
  $\nu_3$ &  & 0.054 & 0.055 & 0.049 & 0.108 & 0.056 & 0.054 \\ 
  $\nu_4$ &  & 0.061 & 0.062 & 0.057 & 0.077 & 0.061 & 0.061 \\ 
  $k_1$ &  & 0.065 & 0.063 & 0.061 & 0.062 & 0.059 & 0.056 \\ 
  $\gamma_1$ &  & 0.061 & 0.057 & 0.069 & 0.015 & 0.064 & 0.061 \\ 
  $\gamma_2$ &  & 0.070 & 0.080 & 0.075 & 0.108 & 0.077 & 0.062 \\ 
  $\lambda_2$ &  & 0.063 & 0.067 & 0.062 & 0.062 & 0.092 & 0.079 \\ 
  $\lambda_3$ &  & 0.059 & 0.065 & 0.063 & 0.246 & 0.100 & 0.086 \\ 
  $\lambda_4$ &  & 0.060 & 0.066 & 0.068 & 0.185 & 0.070 & 0.054 \\ 
   \bottomrule
\end{tabular}
\caption{Comparison of the type 1 error of Wald tests for various estimation methods in Study B under a correctly specified model. 
No small sample correction is used. 
Robust ML corresponds to the use of robust Wald tests.
NA indicates that the estimator never converged in the simulations.
The R packages lava and MIIVsem were used for IV estimation. GLS and WLS were carried out using the R package lavaan.
} 
\label{tab:comparison}
\end{table}

\clearpage

\section{Growth of guinea pigs}
\label{SM:figPig}
\begin{figure}[!h]
\centering
\includegraphics[width=\textwidth]{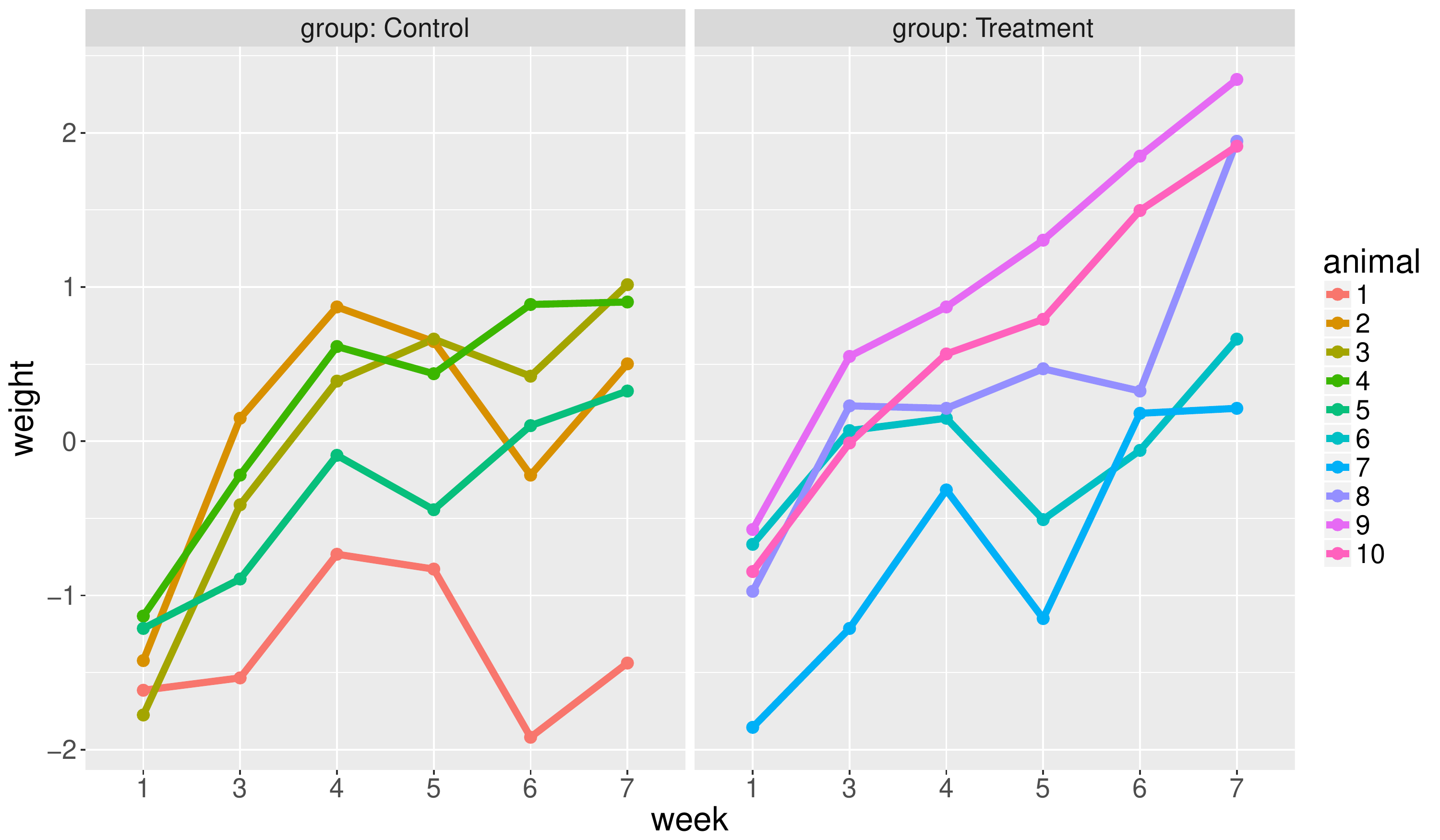}
\caption{\label{fig:spaghetti}
Spaghetti plot of the weight of guinea pigs over weeks.}
\end{figure}

\section*{Reference}
\label{sec:org4cdc50f}
\begingroup
\renewcommand{\section}[2]{}
\bibliographystyle{apalike}
\bibliography{bibliography}
\endgroup